\documentclass[aps,prd,twocolumn,nofootinbib,longbibliography,10pt]{revtex4-2}
\usepackage{bm}
\usepackage{adjustbox}
\usepackage{latexsym}
\usepackage{amsmath,amsfonts,amssymb}
\usepackage{graphicx,epsfig}
\usepackage{psfrag}
\usepackage{amsthm}
\usepackage{nicefrac}
\usepackage{bigints}
\usepackage{xcolor}
\usepackage{amsmath}
\usepackage[colorlinks=true,urlcolor=blue,citecolor=red,linkcolor=blue]{hyperref}
\usepackage{braket}
\usepackage[normalem]{ulem}
\usepackage{multirow}
\useunder{\uline}{\ul}{}
\usepackage{environ}
\usepackage{mathtools}
\usepackage{relsize}
\usepackage{caption}
\usepackage{subcaption}
\usepackage{placeins}
\usepackage[shortlabels]{enumitem}

\setlist[itemize]{itemsep=0.35em, topsep=0.4em, parsep=0.15em}
\setlist[enumerate]{itemsep=0.35em, topsep=0.4em, parsep=0.15em}
\def\s{\sigma} 
\def\o{\omega}

\def\a{\alpha}
\def\b{\beta}

\def\d{\delta}
\def\D{\Delta}
\def\P{\Phi}

\def\t{\tau}

\def\f{\frac}

\newcommand{\be}{\begin{equation}}
\newcommand{\ee}{\end{equation}}
\newcommand{\bes}{\begin{equation*}}
\newcommand{\ees}{\end{equation*}}

\newcommand{\beq}{\begin{eqnarray}}
\newcommand{\eeq}{\end{eqnarray}}

\allowdisplaybreaks

\begin{document}
\title{\textbf{Distinguishing Flat and Warped Extra Dimensions with Unruh-DeWitt Detectors}}
\author{Ashmita Das}
\email{ashmita.phy@gmail.com}
\affiliation{Department of Physics, SRM University AP, Amaravati 522240, India}
\author{Anjana Krishnan}
\email{anjana$_$krishnan@srmap.edu.in}
\affiliation{Department of Physics, SRM University AP, Amaravati 522240, India}
\author{Soumitra Sengupta}
\email{tpssg@iacs.res.in, soumitraiacs@gmail.com}
\affiliation{{School of Physical Sciences,}\\
{Indian Association for the Cultivation of Science,}\\
{2A \& 2B Raja S C Mullick road, Jadavpur, Kolkata 700032, India}}

\begin{abstract}
%%%
\noindent
We examine the response of a two level Unruh-DeWitt (UD) detector interacting with the Kaluza-Klein (KK) mass modes of a bulk scalar field in  extra dimensional models. Considering two of the promising extra dimensional models with distinct compactifications, such as large, flat extra dimensions in the Arkani Hamed, Dimopoulos \& Dvali (ADD) framework and warped extra dimensions in the Randall-Sundrum (RS) scenario, we compute the profile function of the detector. We extend this computation to the 6 dimensional doubly compactified spacetime geometry. The profile function of the detector has been examined as a function of the energy gap of the detector, the compactification radius of the extra dimension, and the dimensionality of spacetime, requiring the chosen parameter space appropriately such that the gauge hierarchy problem is duly addressed. We find that the pattern of the profile function is dictated by the spacing of the KK mass spectrum and the resolution of the detector. 
For ADD model, the spacing between successive KK modes decreases with the increasing compactification radius, yielding a smoothly increasing continuum pattern in the profile function. Whereas for the RS warped geometry model, a tiny deviation in the radius of the extra dimension produces a large change in the same, resulting a sharp increment. 
% For ADD model the KK mass spacing is extremely small due to the large radius requirement, while for RS scenario, the same saturates at a TeV scale as fixed by the warp factor. 
% This difference in the mass gap determines whether individual KK modes remain distinguishable for a given energy resolution of the detector. 
% This offers a geometry dependent signature which differentiates between the two classes of extra dimensional models. 
The resolvability of the KK mode structure is sensitive to the underlying geometry of the spacetime, providing a qualitative distinction between flat and warped extra dimensional scenarios within the UD detector framework.
\end{abstract}
\maketitle
\section{Introduction}\label{intro}
\noindent 
The possibility that spacetime may possess dimensions beyond the familiar four has a long history in theoretical physics. The idea can be traced back to the pioneering works of Kaluza and Klein, who proposed an additional compact spatial dimension in an attempt to unify gravity and electromagnetism within a higher dimensional geometric framework \cite{Kaluza:1921tu,Klein:1926tv}. Although the original Kaluza--Klein (KK) proposal did not provide a complete description of nature, it established a remarkable connection between higher dimensional geometry and the gauge structure of four dimensional physics. The idea of extra dimensions subsequently reappeared in a much broader context with the development of higher dimensional field theories and, most notably, string theory, in which additional spatial dimensions arise naturally as a consequence of the consistency of the theory \cite{Green:1987mn,Green:1987sp,Polchinski:1998rq,Polchinski:1998rr}.
Interest in extra dimensional theories was further renewed by the possibility that the geometry of the extra dimensions might provide an explanation for the apparent hierarchy between the electroweak and Planck scales. In particular, the proposal of large extra dimensions by Arkani-Hamed, Dimopoulos and Dvali (ADD) demonstrated that the fundamental gravitational scale could, in principle, be substantially lower than the four dimensional Planck scale if gravity propagates in additional spatial dimensions \cite{Arkani-Hamed:1998jmv,Antoniadis:1998ig,Arkani-Hamed:1998sfv}. A qualitatively different realization was subsequently proposed by Randall and Sundrum, in which the extra dimension is strongly curved and the corresponding warp factor generates an exponential separation between the fundamental and electroweak scales \cite{Randall:1999ee,Randall:1999vf}. These developments revealed that the phenomenology of extra dimensions is determined not merely by their number and size, but also, crucially, by their geometric structure. In particular, flat and warped extra dimensional scenarios can lead to substantially different KK spectra, localization properties and effective couplings to four dimensional matter \cite{Antoniadis:1998ig,Appelquist:2000nn}.
The possibility of extra dimensions has therefore motivated an extensive programme of phenomenological and observational investigations. At the level of high energy physics, searches for KK excitations, missing energy signatures, virtual graviton exchange, microscopic black holes and other manifestations of higher dimensional dynamics have been pursued at particle colliders, including the Large Hadron Collider \cite{CMS:2014jvv,CMS:2017boz,CMS:2017qyo,CMS:2018nlk,CMS:2024nht,CMS:2025cgw,CMS:2026cgd,ATLAS:2016bek,ATLAS:2017ayi,ATLAS:2021kxv}.
In parallel, extra dimensional scenarios have been investigated through their possible implications for cosmology and astroparticle physics, including the early universe evolution, dark matter, gravitational interactions at short distances, and the production or propagation of higher dimensional modes \cite{Appelquist:2000nn,Dienes:2002hg, Rubakov:2001kp,Csaki:2004ay}. These studies have placed increasingly stringent constraints on various classes of extra dimensional models, although the phenomenological viability and characteristic signatures depend strongly on the geometry of the bulk, the localization of matter fields and the mechanism of compactification. Thus, the search for extra dimensions is not restricted to a particular experimental scale or observational arena; it encompasses particle physics, gravitation, cosmology and astrophysics.
%%%%%%%
A complementary perspective arises from quantum field theory in curved spacetime. The Unruh effect demonstrates that the response of a quantum detector is sensitive to the relation between the quantum state of a field and the spacetime trajectory of the detector \cite{Unruh:1976db}. The Unruh--DeWitt (UD) detector provides a particularly simple and useful idealization of such a probe, consisting of a localized two-level system coupled to a quantum field along its worldline \cite{DeWitt:1980hx}. Its transition probability, or equivalently its response function, is determined by the two-point correlation function of the field evaluated along the detector trajectory. This formulation has subsequently been developed extensively in flat and curved spacetime and provides a well-defined framework for studying the response of localized quantum probes to nontrivial gravitational backgrounds \cite{Louko:2007mu}. 
%%%%%%
In this work, we explore whether such a detector can provide a new diagnostic of extra dimensional geometry. Rather than searching directly for KK particles at high energies or for modifications of gravitational interactions, we ask whether the response of a quantum detector can encode information about the structure of the extra dimension. In particular, we consider the response of an UD detector in different higher dimensional backgrounds and investigate whether the resulting detector spectrum can distinguish between a flat extra dimension and a warped extra dimension. The underlying motivation is that the detector does not probe the bulk geometry through a single effective parameter; its response is determined by the quantum field correlations, which in turn contain information about the spectrum and wave functions of the higher dimensional modes.
This provides a potentially useful and conceptually distinct approach to the problem of identifying the geometry of extra dimensions. A flat and a warped extra dimension can give rise to qualitatively different mode structures and localization properties of the KK mode wave functions. Since these differences enter the Wightman function and consequently the transition probability of a localized detector, the detector response may retain information about the underlying bulk geometry that is not immediately apparent from the four dimensional effective theory. The purpose of the present work is therefore to investigate the extent to which the UD detector can serve as a probe of extra dimensional geometry, with particular emphasis on distinguishing flat and warped extra dimensional scenarios.
In summary, the novelty of our work 
is that the detector response samples the two-point correlation function of the bulk scalar field and hence the KK mode structure, potentially allowing the geometry, either flat/ warped, to leave a distinctive imprint.

\noindent
We present our manuscript in the following order. 
In sec.(\ref{res_ud_1}), we briefly describe the derivation of the profile function of the UD detector while interacting with a single massive scalar field. In the secs. (\ref{flat_spectrum_1}) and (\ref{rs_spectrum_1}), we describe the emergence of KK mass modes for a bulk scalar field and derive the profile function of the UD detector due to the interaction with the infinite number of KK mass modes in the background of flat and warped extra dimensional models. Further we 
analyze the parametric variations of the profile function in Secs. (\ref{plot_add}) and (\ref{plot_rs}) for both the extra dimensional scenarios. We discuss our results in Sec. (\ref{discussion}). For the self consistency of the manuscript, we attach appendices in the App. (\ref{appendix_2}) - (\ref{rs_appendix_1}). 
Throughout the manuscript, we follow the $\{-,+,...,+\}$ metric signature. 
%%%%
%%%
\section{Response of UD detector: transition probabilities and rates}\label{res_ud_1}
In this section we revisit the mechanism of UD detector and derive the transition probabilities associated with it. A UD detector can be portrayed as a two level atom with energy eigenstates represented as $\ket{E_1}$ and $\ket{E_2}$, with eigenvalues $E_1<E_2$. 
 One can formulate the interaction Hamiltonian of a UD detector interacting with a scalar field $\phi(x(\tau))$ of mass $m$
as follows,
%%%
\begin{equation}
    H_{\rm int} = g \mu(\tau) \phi(x(\tau))\,.\label{int_1_st}
\end{equation}
%%%
Here $g$ represents the coupling constant and $\mu(\tau)=\exp{(iH_D\tau)}\mu(0)\exp{(-iH_D\tau)}$ is the detector's monopole moment in the interaction picture. $H_D$ denotes the Hamiltonian for the UD detector.
%%%%
Decomposing the field mode in Minkowski quantization, we write, 
\[
\phi(x(\tau)) = \int \frac{d^3 p}{\sqrt{2(2\pi)^3\omega_p}}(u_p a_p + u_p^*a_p^\dagger)\,,
\]
where $a_p,\,a_p^\dagger$ are the annihilation and creation operators with the field modes written as, $u_p= e^{-i(\omega_{\bar{p}}t-\bar{p}.\bar{x})}$. Here $\o_{\bar{p}}\equiv \sqrt{|\bar{p}|^2+m^2}$.
%%
%%%%%%
%%%%%%
The detector-field interaction gives rise to several possible transitions between its initial and final states, which are classified as below:
\\
\noindent
%\begin{itemize}
1. Total and spontaneous emission: The total emission rate is the sum of the spontaneous and induced emission rates of the system. The total and the spontaneous emission rates are calculated on the same worldline of the detector, while the initial states of the matter field are different. 
The initial and final states for the total transition and the spontaneous emission of the system can be depicted as, $\left(\ket{E_2,0_M}\rightarrow\ket{E_1,1_p}\right)$, where $\ket{0_M},\,\ket{1_p}$ are the vacuum and the 1-particle states for the scalar field in Minkowski spacetime and $\left(\ket{E_2,0_R}\rightarrow\ket{E_1,1_{k_\perp}}\right)$, where $\ket{0_R}$, $\ket{1_{k_\perp}}$ represent vacuum and 1-particle state of the field in Rindler spacetime \cite{Agullo:2010iq} respectively. Note that for both the processes the detector follows a uniformly accelerated trajectory and thus, in the process of spontaneous emission, the field lies in the vacuum state associated with the detector's trajectory. Therefore, any difference between these two responses 
stems from the initial states of the field which trigger induced emission in the system.
\footnote{The spontaneous emission rate is independent of the detector's trajectory for a massless scalar field \cite{Agullo:2010iq}.  Therefore, this process can also be realized with respect to the inertial observer in the Minkowski spacetime for the transition of the system such as: $\left(\ket{E_2,0_M}\rightarrow\ket{E_1,1_p}\right)$ for a massless scalar field. It's noteworthy that for a massive scalar field, this is not strictly followed.}
\\
\noindent
2. Induced emission: As mentioned earlier, the induced emission rate of a system is a numerical difference between the total and spontaneous emission rates. For the 
induced emission, the initial state of the field ($\ket{0_M}$) already contains background field quanta from the perspective of a uniformly accelerated detector. Thus, the transition rate is enhanced beyond the spontaneous emission rate.
%%%
\\
\noindent
3. Absorption: In this process, the UD detector gets excited and makes a transition from $\ket{E_1} \,\to\,\ket{E_2}$ along with the emission of a field quantum. The initial and final states of the system can be written as, 
$\left(\ket{E_1,0_M}\rightarrow\ket{E_2,1_p}\right)$.
%\end{itemize}
%%%%
In general, the transition rate of such a system is obtained via first-order perturbation theory, beginning with the transition amplitude, 
\begin{align}
    A_{i\to f}&= -\int d\tau \braket{E_f,1_p|g\,\mu(\tau)\phi(x(\tau))|E_i,0_M}\,,
  \label{eq: Transition amplitude}
  \end{align}
  %%%%%
  which leads to the transition probability as follows, 
  \begin{align}
      P_{i\to f}= \int d^3p|A_{i\to f}|^2 = CF(\Delta E)\,,
  \end{align}
where $C= g^2|\braket{E_f|\mu(0)|E_i}|^2$, $\D E=E_f-E_i$, and 
\begin{align}
    F(\Delta E) &= \int d\tau \,d\tau^\prime e^{-i\Delta E(\tau-\tau')}\braket{0_M|\phi(x)\phi(x')|0_M}\notag\\
    &=\int d\tau\,d\tau^\prime e^{-i\Delta E(\tau-\tau')}G_+(\tau,\tau')\,,
    \label{eq:res_fun_gen_1}
\end{align}
known to be the response function of the detector. In order to obtain the total emission rate (spontaneous + induced) of the system due to the detector-field interaction, we consider $E_f=E_1$ and $E_i=E_2$ \cite{Alkofer:2016utc}. 
%On the other hand absorption process demands that $E_f=E_2$ and $E_i=E_1$ \cite{Alkofer:2016utc}. 
%%%
In the above expression, $G_+(x,x')=\braket{0_M|\phi(x')\phi(x)|0_M}$ represents the positive frequency Wightman function which becomes,  
\begin{align}
    G_+ (x, x')&=\int \frac{d^3\bar{p}}{(2\pi)^3}\frac{e^{i(\bar{p}.\Delta \bar{x}-\omega_{\bar{p}} \Delta t)}}{2\omega_{\bar{p}}}\notag\\
    &=\int \frac{d|\bar{p}|}{(2\pi)^2}\frac{|\bar{p}|^2\,e^{i(\bar{p}.\Delta \bar{x}-\omega_{\bar{p}} \Delta t)}}{\omega_{\bar{p}}}\label{eq: Wightman function}
\end{align}
%%%
\footnote{By definition, $\tilde{G}(p^2)=(p^2-m^2)^{-1}$. Thus the Wightman function in position space can also be written as, $G_+(\bar{x}, t)=-i \int \frac{d^3\bar{p}\,dp_0}{(2\pi)^4}\tilde{G}(p^2)e^{i(\bar{p}.\Delta\bar{x}-p_0\Delta t})$\label{footnote_1}}
%%%
In Eq. (\ref{eq:res_fun_gen_1}), we express $\bar{\tau}=\tau+\tau'/2$ and $\Delta\tau=\tau-\tau'$, and obtain, 
%%%%%
\begin{align}
F_{\rm TE}(\Delta E)= \int d\bar{\tau}\,d\Delta\tau e^{-i
\Delta E\, \Delta \tau} G_+(\Delta\tau)\,.
\label{response_11}
\end{align}
Note that in Eq. (\ref{response_11}), $F_{\rm TE}$ denotes the response function of the detector for the total emission probability includes both the spontaneous and induced emission processes \cite{Alkofer:2016utc}. 
\\
\noindent
We consider a finite time interval of the interaction between the detector and the field, spanning from $-\t_0$ to $+\t_0$. Integrating with respect to $\bar{\t}$, we obtain an overall factor of $2\t_0$ on the right-hand side of Eq. (\ref{response_11}). Dividing  $F(\D E)$ by this factor $2\t_0$, and subsequently taking the interaction time to infinity (corresponds to the detector-field interaction that is switched on and off infinitely slowly), we obtain the response rate for the total emission as follows, 
% \textcolor{red}{include these steps in appendix calc}, 
%%%
\begin{align}
    \dot{F}_{\rm TE}(\Delta E)= \int_{-\infty}^{\infty} d\Delta\tau\, e^{-i\,\Delta E\,\Delta \tau}\,\,G_+(\Delta\tau- i \epsilon)\,.
    \label{trans_prob_rate_1}
\end{align}
In the above equation, the $i\epsilon$ prescription is introduced such that $\D\t \to (\D \t - i \epsilon)$, $\epsilon \to 0^{+}$, in order to regulate the two-point correlation function and ensure its convergence. 
We refer our readers to the Appendix (\ref{appendix_2}) for the detailed calculation of the transition rate. 
%%%%
\subsection{Thermality of the response function: the KMS condition}\label{kms}
%%%%%
We now establish the thermal behaviour of the two-point correlation function when evaluated on the accelerated trajectory, and the Kubo-Martin-Schwinger (KMS) condition. The vacuum state of a scalar field $\ket{0_M}$ is Lorentz invariant, which restricts the two-point correlation function $G_{+}(x,x')$ to depend on the spacetime interval {\it i.e}, $\s=(t-t')^2-(x-x')^2$, which is a Lorentz invariant quantity. Using the accelerated trajectory of the detector with uniform acceleration $a$ as below, 
\beq
t=\f{1}{a}\,{\rm sinh}(a\t)\,,\,\,\,\,\,\,x=\f{1}{a}\,{\rm cosh}(a\t)\,\,\,\,\,{\rm and}\,\,\,\,\,\,y=z=0
\label{rind_traj}
\eeq
we obtain the spacetime interval in terms of the Rindler proper time as, 
\beq
\s=\,\f{2}{a^2}\left[{\rm cosh}(a\D \t)-1\right]
\label{spt_int_1}
\eeq
It can be noted that while restricted to the above worldline, $G_{+}$ depends only on $\D \t$ via ${\rm cosh}(a\D \t)$. Using the identity ${\rm cosh}(x+ i y)=\left({\rm cosh} x\,{\rm cos} y+ i\, {\rm sinh} x\,{\rm sin} y\right) $ and performing a shift such as $\D \t \to \D \t + 2\pi\,i/a$, it can be shown that $G_{+}\left(\D \t + 2\pi\,i/a\right)=G_{+}(\D \t)$, which demonstrates that the Wightman function is periodic in imaginary time with period $\b=2\pi/a$. This is known as the KMS condition on the two-point correlation function. The periodicity relation as established is precisely the characteristic property of a two-point correlation function evaluated in a thermal state with temperature $T=1/\b=a/2\pi$ \cite{Unruh:1976db}. Thus, when the Minkowski two-point correlation function is restricted to the worldline of a uniformly accelerated observer, the Minkowski vacuum is perceived as a thermal state with temperature $T=a/2\pi$, without requiring any prior thermal state as a reference. 

\noindent
Regarding the analytic property of the $G_{+}(\D \t)$, we discuss several aspects as below: 
\begin{itemize}
\item
Note that for $\s\to 0$ {\it i.e} $\D \t=0$, $G_{+}(\D \t)$ acquires singularity. Subsequently, the periodicity condition of the two-point correlation function demands that the singularity recurs at every point where $\D \t=i n \b$ and $n \in \mathbb{Z}$. 
\item 
It is already perceived that the convergence of $G_{+}(\D \t)$ requires a negative shift in the imaginary $\D \t$ axis. This implies that $G_{+}(\D \t)$ is well defined throughout the lower half-plane of the imaginary $\D \t$ axis, such that ${\rm Im}\,(\D \t) < 0 $. 
\item 
We found that  $G_{+}(\D \t)$ has singularities at $\D \t=i n \b$, $n \in \mathbb{Z}$. One can identify the strip $-\b < {\rm Im}(\D \t)<0$, bounded by the two singularities nearest to the real $\D \t$ axis, such as $n=-1$ to $n=0$. Within this region, $-\b < {\rm Im}(\D \t)<0$, right below the real $\D \t$ axis, $G_{+}(\D \t- i \epsilon)$, $\epsilon \to 0^{+}$, is well defined and analytic.
\end{itemize}
In the context of detector-field interaction, the KMS condition can equivalently be portrayed in terms of the response function, as response function is the Fourier transformation of $G_{+}(\D \t)$. Following \cite{Alkofer:2016utc}, one can show that, 
\begin{align}
    \dot{F}(\D E) = e^{-\beta\Delta E}\dot{F}(- \D E)\,,
    \label{kms_2}
\end{align}
where $\beta=1/T$, $T$ being the temperature of the thermal bath. Eq. (\ref{kms_2}) is an equivalent form of the KMS condition in terms of the detector response function. For the above proof, we refer our readers to the Appendix (\ref{periodicity_appendix}).
\\
\noindent
It can be seen that the response rate for the absorption process is obtained by replacing the parameter $\D E$ with $- \,\D E$ in Eq. (\ref{trans_prob_rate_1}), yielding $\dot{F}_{\rm TE}(\D E)=\dot{F}_A(-\D E)$. Furthermore, as discussed earlier, the total emission rate can be expressed as the sum of the induced ($\dot{F}_I(\D E)$) and spontaneous ($\dot{F}_S(\D E)$) emission rates, 
\beq
\dot{F}_{\rm TE}(\D E)=\dot{F}_I (\D E)+\dot{F}_S (\D E)\label{tot_em_1}
\eeq
%%%
Using the KMS condition in Eq. (\ref{kms_2}), one writes, 
\beq
    \dot{F}_A(\D E)&=&e^{-\beta \D E}\dot{F}_A(- \D E)\nonumber\\
    &=&e^{-\beta \D E}(\dot{F}_I(\D E)+\dot{F}_S(\D E))
    \label{eq: induced response rate}
\eeq
%%%%%
If it is assumed that 
\beq
\dot{F}_A(\D E)\approx\dot{F}_I(\D E)\,,
\label{assump_1}
\eeq
one obtains, 
\beq
\dot{F}_I(\D E) = \frac{\dot{F}_S(\D E)}{e^{\beta \D E}-1}
\label{induced_em_12}
\eeq
%%%
As discussed earlier, the spontaneous emission rate is independent of the detector's trajectory for a massless scalar field. Consequently, Eq. (\ref{induced_em_12}) is exactly satisfied for both the inertial observers in Rindler/ Minkowski spacetime 
%(As $\dot{F}_S(\D E)$ is the same for both the observers). 
This leads to comment that 
the assumption in Eq. (\ref{assump_1}) is also valid for the massless case. 
%%%
%%%
However, for a massive scalar field, the spontaneous emission rates evaluated for an inertial observer in Minkowski and Rindler spacetime do not precisely coincide \cite{Alkofer:2016utc}. Instead, they coincide only within a restricted region of parameter space and hence the validity of Eqs. (\ref{induced_em_12}) and (\ref{assump_1}) are expected to hold only within the same region of parameter space.  
%%%%%
In the following section, we briefly discuss the distinctive features of detector response for the massive and massless scalar fields. 
%%%
%%
\subsection{Spontaneous emission for a massive scalar field: Minkowski vs Rindler spacetime}\label{spont_massive}
%%%%
We compute the spontaneous emission rate using the simplest possible case, specifically, an inertial observer at rest in Minkowski spacetime. Such an observer follows the trajectory, $\Delta t=\Delta \tau=\tau$, $\Delta x = 0$, where $\tau$ denotes the proper time elapsed along the worldline of the detector while $\D x$ denotes the change in its spatial position. We consider that the detector is interacting with a massive scalar field of mass $m$, which is in the Minkowski vacuum state $\ket{0_M}$.  Therefore, following the derivation of Eq. (\ref{trans_prob_rate_1}), the spontaneous emission rate turns out to be
%%%
\beq
    \dot{F}_s(\D E)&=&\frac{1}{2\pi}\,\int \frac{dp\,p^{2}}{\sqrt{\left|\bar{p}\right|^2+m^2}}\delta\left(\D E-\sqrt{\left|\bar{p}\right|^2+m^2}\right)\nonumber\\
    &=&\,
     \frac{1}{2\pi}\,\sqrt{(\D E)^2-m^2}\,\,\Theta(\D E-m)
     \label{spon_em_1}
\eeq
%%%%
%%%%
The spontaneous emission rate can also be computed by implementing Rindler quantization, where the massive scalar field $\phi(x)$ can be decomposed as follows \cite{Alkofer:2016utc}, 
\beq
\phi(x)=\int d^3 k\,\left(b_{\tilde{\o},\bar{k}_{\perp}}\,v_{\tilde{\o},\bar{k}_{\perp}}+\,b_{\tilde{\o},\bar{k}_{\perp}}^{\dagger}\,v_{\tilde{\o},\bar{k}_{\perp}}^{*}\right)\,.
\label{rind_quant_1}
\eeq
Here the position space Rindler basis is $\left(\t,\,\xi,\,x_{\perp}\right)$ and $\bar{k}_{\perp}=(k_y,\,k_z)$. $\tilde{\o}$ depicts the field frequency. The Rindler mode function can be expressed as \cite{Takagi:1986kn, Alkofer:2016utc}, 
\begin{equation}
v_{\tilde{\o},\bar{k}_{\perp}}=\left[\f{{\rm sinh}\left(\pi \tilde{\o}/a\right)}{4 \pi^4\,a}\right]^{1/2}\,K_{i \tilde{\o}/a}\left(\f{\sqrt{|\bar{k}_{\perp}|^2+m^2}}{a}\,e^{a\xi}\right)e^{-i\omega\tau}\,,
\label{mode_rindler_1}
\end{equation}
% from Crispino eq. 2.92
where $a$ is the acceleration of the detector.
%%%
Using Rindler quantization and the proper frame of the Rindler observer as $\xi=0$ and $x_{\perp}=0$, the spontaneous emission rate in Rindler spacetime turns out to be \cite{Agullo:2010iq,Alkofer:2016utc},  
\begin{align}
\dot{F}_{S}(\D E)=2\pi \int & d^2k_{\perp}\,d\tilde{\o}\,\left|K_{i \tilde{\o}/a}\left(\f{\sqrt{|\bar{k}_{\perp}|^2+m^2}}{a}\right)\right|^2\notag\\
&\times\,\f{{\rm sinh}(\pi \tilde{\o}/a)}{4 \pi^4\,a}\,\d(\tilde{\o}-\D E)
\label{res_rindler_spon}
\end{align}
%%%%
For the derivation of Eq. (\ref{res_rindler_spon}), we recommend readers to Secs. (2.3) of \cite{Alkofer:2016utc}, \cite{Agullo:2010iq} and Appendix (\ref{spont_ems_1}) of the present manuscript.
It can be noted that Eqs. (\ref{spon_em_1}) and (\ref{res_rindler_spon}) do not exactly coincide. A clear difference can be identified by the presence of the term $\Theta(\D E - m)$ in Eq.(\ref{spon_em_1}). The spontaneous emission rate corresponding to an inertial observer in Minkowski spacetime reduces exactly to zero under the condition $E<m$. Whereas Eq. (\ref{res_rindler_spon}), remains small but nonzero for $E<m$ (see Fig. (1) of \cite{Alkofer:2016utc} for a numerical depiction and the relevant discussion therein). \cite{Alkofer:2016utc}, asserts that Eqs. (\ref{spon_em_1}) and (\ref{res_rindler_spon}) coincide asymptotically for $E\gg m$ and consequently, in this regime the Eq. (\ref{induced_em_12}) and (\ref{assump_1}) can be trusted. 
Therefore, under this condition, using Eq. (\ref{induced_em_12}), the response rates for the induced emission and absorption processes turn out to be as follows, 
\beq
\dot{F}_{I}(\D E)&=&\dot{F}_{A}(\D E)=\f{1}{2\pi}\f{\sqrt{(\D E)^2-m^2}\,\,\Theta(\D E -m)}{e^{\b\,\D E}-1}\nonumber\\
&=&\f{1}{2\pi}\,\,\f{\mathcal{F}(\D E)}{e^{\b\,\D E}-1}\,,
\label{abs_final_1}
\eeq
%%%
where, 
\beq
\mathcal{F}(\D E)=\sqrt{(\D E)^2-m^2}\,\,\Theta(\D E -m)\,,
\label{profile_func_1}
\eeq
known as the profile function \cite{Alkofer:2016utc}. In the subsequent sections, we explore the profile function of the UD detector while it is interacting with the KK modes of a bulk scalar field in the background of flat and warped extra dimensional models.   
%%%%
%%%%
\section{Profile function in Large and flat extra dimensional model}\label{flat_spectrum_1}
%%%
In 1920, Kaluza and Klein (KK) introduced the theory of extra dimensions with a motivation to unify the fundamental interactions of nature. According to the KK theory, at every point of our 4 dimensional noncompact spacetime, there exists an extra spatial compactified dimension, resulting the spacetime to be $5$ dimensional. Considering a circular compactification the spacetime topology becomes $\mathcal{R}_4\times \mathcal{S}_1$, which govern by the spacetime metric: $ds^2=g_{MN}\,dx^{M}\,dx^{N}$, where ${M,\,N=0,..4}$ with the metric signature is $(-,+,..,+)$. 
% \beq
% ds^2=\eta_{\mu\nu} dx^{\mu}\,dx^{\nu}+R^2 dy^2\label{flat_metric_1}\,. 
% \eeq
%%%
$R$ is the radius of the circular compactification and $y$ represents the angular coordinate for the extra dimension, satisfying the periodic boundary condition $y\to y+2\pi R$. 
%%%%
%%%%
Advancing KK theory, Arkani-Hamed et al., have shown in \cite{Arkani-Hamed:1998jmv} that large and flat extra dimensional models resolve the gauge-hierarchy problem, which arises due to the large gap between the two fundamental energy scales such as Planck $(\sim 10^{16}\,{\rm TeV})$ TeV and electroweak scale $(\sim 1\,{\rm TeV})$. In literature, this framework is popularly known as ADD model \cite{Arkani-Hamed:1998jmv}, where a $\mathcal{R}_4\times \mathcal{M}_d$, $d\geq 2$ dimensional spacetime is considered. Here $\mathcal{M}_d$ represents a $d$ dimensional compact manifold of volume $\sim R^d$, where, $R\sim 10^{32/d}\,{\rm TeV}^{-1}$. To estimate the value of $R$, one uses $M^{2}_{pl}=M_{*}^{d+2}\,(2\pi R)^d$, where $M_{pl} \sim \mathcal{O}\left(10^{16}\right)$ TeV and the fundamental Planck scale $M_{*} \sim \mathcal{O}\left(1\right)$ TeV in order to resembles with the electroweak scale in our $(3+1)$ dimensional Universe.  This implies that the strength of the gravitational interaction becomes comparable to that of the electroweak interactions, thereby suggesting a viable solution to the gauge-hierarchy problem. 
%%%
Gravity appears to be weak in the effective $(3+1)$ dimensional theory because it propagates throughout the $(4+d)$ dimensional bulk and is effectively diluted over the large extra dimensional volume  $\sim R^{d}$. On the contrary, standard model fields are confined to a $(3+1)$ dimensional submanifold characterized by the electroweak length scale.
In this manuscript, we first consider a bulk scalar field ($\Phi(x^{\mu}, y)$) with  mass $m_0$, propagating in a 5 dimensional ADD spacetime. For a circular compactification, the field satisfies the periodic boundary condition $y \to y+2\pi R$ and can be expressed as, 
%%%%%
\begin{equation}
\Phi(x^\mu,y)=\frac{1}{\sqrt{2\pi R}}\sum_{n=-\infty}^\infty\phi_n(x^\mu)e^{i n y/R}\,.
\label{kk_mode_1}
\end{equation}
We consider a 5 dimensional action for a scalar field of mass $m_0$, as below, 
%%%%
\beq
S_5=\int d^4x \int_{0}^{2 \pi R} dy\,\left[-\,\f{1}{2}\partial_{M}\Phi\,\partial^{M}\Phi-\f{1}{2}m_{0}^{2}\,\Phi^2\right]\,.\label{scalar_1}
\eeq
%%%
Using Eq. (\ref{kk_mode_1}) in the above action and integrating with respect to $y$, one obtains the 4 dimensional effective action for the scalar filed as follows, 
\beq
S_{4}=\sum_{n=0}^{\infty}\int d^4x\,\left[-\,\f{1}{2}\partial_{\mu}\phi_n\,\partial^{\mu}\phi_n\,-\f{1}{2}\left(m_{0}^{2}+\f{n^2}{R^2}\right)\phi_{n}^{2}\right].
\label{scalar_4d_1}
\eeq
%%%
Thus KK mass tower has been generated in the ($3+1$) dimensional effective theory with $m_n^2= m_0^2+n^2/R^2$. 
%%%%%
The 4 dimensional Klein-Gordon (KG) equation for the action in Eq. (\ref{scalar_4d_1}) becomes, 
\beq
\sum_{n=0}^{\infty}\,\left(\Box -\,m_{n}^{2}\right)\,\phi_n=0 \nonumber
\eeq
where upon quantization, the field solution turns out to be, 
\beq
\phi_{n}(x) = \sum_{n=0}^{\infty}\int \frac{d^3\bar{p}}
{\sqrt{2(2\pi)^3\omega_{\bar{p}}}}\,\left[e^{-i(\omega_{\bar{p}}-\bar{p}.\bar{x})} a_p +\, {\rm H.C}\right]\,.\label{field_sol_1}
\eeq
%%%%
We consider a two-level UD detector, interacting with the KK mass tower of a scalar field on the $(3+1)$ dimensional submanifold. In order to compute the profile function of the detector, we follow the discussion as developed in Sec. (\ref{res_ud_1}) and write the interaction Hamiltonian as follows,
%%%%
\beq
 H = g \,\mu(\tau) \phi_{n}(x(\tau))\,,
 \label{hamil_add_1}
\eeq
%%%
where $g,\,\t,\,\mu(\t)$ bear the same notions as mentioned in the Sec. (\ref{res_ud_1}) and $\phi_{n}(x)$ is depicted in Eq. (\ref{field_sol_1}).   Thus the generic form of the transition rate of the detector becomes,
%\cite{Alkofer:2016utc, Agullo:2010iq}, 
%%%%
\beq
\dot{F}(\D E)&=&\,\sum_{n=0}^{\infty}\,C\int d(\D\tau)\,e^{-i E\Delta\tau}\braket{0_M|\phi_{n}(x')\phi_{n}(x)|0_M}\nonumber\\
&=&\sum_{n=0}^{\infty}\,C\int d(\D\tau)\,e^{-i E\Delta\tau}\,G_{+}^{(n)}(x',x)\label{trans_1}
\eeq
%%%%%
From Eq. (\ref{trans_1}) the effective $(3+1)$ dimensional two-point function can be extracted as, $\sum_n G_{+}^{(n)}(x', x)$. It can be noted that Eq. (\ref{trans_prob_rate_1}) is linear in the Wightman function, which admits that the sum in Eq. (\ref{trans_1}) can equivalently be written in momentum space, yielding the propagator with an infinite tower of simple poles as follows,
%%%%
\begin{equation}
   \tilde{G}(p^2)= \sum_{n=0}^{\infty}\frac{1}{p^2-m_{n}^{2}}\,.
   \label{prop_1}
\end{equation}
%%%%%%
Correspondingly, it can be perceived that the rate of the transition probability of the detector becomes a superposition of the responses associated with all KK masses.

In subsequent analysis, we follow the treatment as developed in \cite{Alkofer:2016utc} to derive the response function of the detector while it interacts with the KK modes. 
We start with a generalized propagator of the form $\tilde{G}(p^2)=\f{1}{P_n (p^2)}$, where  $P_n (p^2)$ is a polynomial in $p^2$. Taking $\mu_i$ (${i=1,2,....,n}$) to denote the $n$ number of roots of $P_n (p^2)$, which determine the pole locations of the propagator, the polynomial can be expressed in the form, 
\beq
P_n (p^2)=\bm{c}\prod_{i=1}^n(p^2-\mu_i)\,,\label{ostro_1}
\eeq
where $\bm{c}$ is the normalization constant. From Eq. (\ref{ostro_1}), one obtains, $\left[P_{n}(p^2)\right]^{-1}=\left[\bm{c}\prod_{i=1}^n(p^2-\mu_i)\right]^{-1}$ which can be rewritten 
as a sum of simple fractions as follows, 
\begin{equation}
\left[P_n(p^2)\right]^{-1}=\frac{1}{\bm{c}}\sum_{i=1}^n\frac{A_i}{(p^2-\mu_i)}\,.\label{eq:polynomial wightman}
\end{equation}
%%%
In Eq. (\ref{eq:polynomial wightman}), $A_i$ measures the strength of the $i$-th propagating mode (or in other words these coefficients $A_i$ characterize the relative contribution associated with each pole $\mu_i$). 
%%%%
In order to determine the coefficients $A_i$, we multiply both side of the Eq. (\ref{eq:polynomial wightman}) by $\prod_{k=1}^{n} (p^2-\mu_k)$, which yields 
% [see appendix for detailed derivation], 
%%%
\begin{align}
    1=\sum_{i=1}^n\,A_i\,\prod_{j\not=i}(p^2-\mu_j)\label{ostro_2}
\end{align}
%%%
To obtain a particular coefficient, one evaluates the above expression at the pole location $p^2=\mu_k$. It can be perceived that 
all terms in the summation vanish except with $i=k$, which leads to, 
\begin{equation}
    A_i= \left(\prod_{j\not=i}\mu_i-\mu_j\right)^{-1}
    \label{ostro_3}
\end{equation}
Its noteworthy that the above decomposition is possible under the assumption that the propagator contains only poles of order 1. This ensures that every factor appears only once in the denominator of Eq. (\ref{ostro_3}) such that $\mu_i \neq \mu_j$ for $i \neq j$. 
%%%
Comparing Eqs. (\ref{prop_1}) and (\ref{ostro_1}), one can write that $\mu_i=m^2$ which denotes square of the mass of the particle. 
At this stage we discuss, some physical restrictions on the roots. For $p^2=\mu_i$, we write from the dispersion relation, $p_{0}=\,\pm\,\left(|\vec{p}|^2+\mu_i\right)^{1/2}$. Since the propagator poles are identified with the corresponding masses of the particles, positive real values of $\mu_i$ imply $p^2=\mu_i=m^2>0$. This condition implies $p_0$ to be real, which in turn yields $e^{-ip_0 t}$ as an oscillatory function that describes stable propagating mode solution for the field. On the other hand the negative real values for $\mu_i$ implies the dispersion relation to be of the form $p_{0}=\,\pm\,\left(|\vec{p}|^2-|\mu_i|\right)^{1/2}$. For $|\vec{p}|<|\mu_i|$, one obtains, $p_0$ to be imaginary. 
This converts $e^{-ip_0 t}\to\,e^{p_0 t}$, a real-valued diverging function in the asymptotic past and future, making it inappropriate to be the viable mode solution for the field. Therefore, we confined ourselves within the real positive values of $p_0$. 

It has already been pointed out that the transition rate of the detector is a linear functional of $G_{+}(x', x)$, and hence $\tilde{G}(p^2)$. Furthermore Eq. (\ref{eq:polynomial wightman}) suggests that $\tilde{G}(p^2)$ can be decomposed into individual terms of the form $\sim \f{A_i}{p^2-\mu_i}$, where each of this term has the form $A_i$ times the standard massive scalar propagator with ${\rm mass}^2=\mu_i$. Combining these two facts, the total transition rate of the detector corresponding to Eq. (\ref{eq:polynomial wightman}) is obtained as the same weighted sum of the transition rate due to single massive scalar field as below,
%%%
\beq
\dot{F}(\D E)=\,\sum_{i}A_i\,\dot{F}^{(0)} (\D E; \mu_i)\,,
\label{gen_trans_rate_1}
\eeq
where $A_i$ is depicted in Eq. (\ref{ostro_3}). 
Following the discussion of the spontaneous, induced, and absorption processes in Secs. (\ref{res_ud_1}), (\ref{kms}), and (\ref{spont_massive}), and using Eqs. (\ref{assump_1}) and (\ref{induced_em_12}), we obtain the induced and hence the absorption rates as follows, 
\beq
\dot{F}_{I}^{(0)}(\D E; \mu)&=&\dot{F}_{A}^{(0)} (\D E; \mu) \nonumber\\
&=& \f{1}{2 \pi} \,\f{\sqrt{(\D E)^2-\mu}\,\,\Theta(\D E-\sqrt{\mu})}{e^{2 \pi \D E/a}-1}\,.
\label{gen_trans_rate_0}
\eeq
Extracting the common thermal factor which is independent of $i$ and comparing with the Eq. (\ref{abs_final_1}), we obtain the profile function for this case to be, 
%%%%
\beq
\mathcal{F}(\D E)=\,\sum_{i}A_i\,\sqrt{(\D E)^2-\mu_i}\,\,\Theta(\D E-\sqrt{\mu_i})
\label{gen_pro_fun_1}
\eeq
This is the general master formula for the profile function of the UD detector, valid for any propagator of the form as written in the Eq. (\ref{eq:polynomial wightman}). Note that the KK propagator in Eq. (\ref{prop_1}) is already in this form with the residue $A_i=A_n=1$ for all $n$ and $\mu_i=\mu_n=m_{n}^{2}=\f{n^2}{R^2}$. Therefore, following the previous discussion for the propagator in Eq. (\ref{eq:polynomial wightman}), one obtains the profile function of the UD detector due to its interaction with the KK mass tower of a massless scalar field in the bulk as follows,
\beq
\mathcal{F}(\D E)&=&\sum_{n=0}^{\infty}\,\sqrt{(\D E)^2- m_{n}^{2}}\,\,\Theta(\D E-m_n)\nonumber\\
&=&\sum_{n=0}^{\infty}\,\sqrt{(\D E)^2- \f{n^{2}}{R^2}}\,\,\Theta\left(\D E-\f{n}{R}\right)
\label{kk_pro_fun_1}
\eeq
Consequently, the emission and absorption rates of the same become, 
%%%%
\begin{align}
\dot{F}_I(\Delta E) &= \dot{F}_A(\Delta E) \notag \\
&= \frac{1}{e^{2\pi \Delta E/a} - 1} \sum_{n=0}^{\infty} \left[ \sqrt{(\Delta E)^2 - \frac{n^2}{R^2}} \; \Theta\left(\Delta E - \frac{n}{R}\right) \right]
\label{kk_trans_rate_1}
\end{align}
%%%
Eq. (\ref{kk_pro_fun_1}) represents the master equation to determine the profile function of the UD detector interacting with the KK mass modes of a massless bulk scalar field in the background of a flat and large extra dimensional model. 

\noindent
Extending the ADD scenario to 6 dimensional spacetime, we obtain the profile function of the detector interacting with the KK modes of a massless bulk field as below, 
%%%
\begin{equation}
    \mathcal{F}(\D E)=\sum_{n=0}^\infty\sum_{q=0}^\infty\sqrt{(\D E)^2-m_{nq}^2}\,\,\Theta(\D E-m_{nq})\label{profile_function_add_1}
\end{equation}
where, in general, $m_{nq}^2=m_0^2+(n^2+q^2)/R^2$ depict the KK mass modes on the $(3+1)$ dimensional submanifold (as we consider a massless bulk scalar field, $m_0$ is not present in the profile function/ rate expressions). For simplicity, we keep $R$, the radii of the compactified spatial extra dimensions same for both the extra dimensions. For the details of the derivation of the profile function for 6 dimensional case, we refer our readers to Appendix (\ref{add_appendix_1}). 
%%%
\section{Profile function in RS warped geometry model}\label{rs_spectrum_1}
%%%% 
One of the most compelling extra dimensional scenarios is the RS model, which dictates a 5 dimensional warped spacetime geometry as follows, 
\beq
ds^2=e^{-2kR_y y}\,\eta_{\mu\nu}\,dx^{\mu}dx^{\nu}+R_{y}^{2}\, dy^2
\label{rs_metric}
\eeq
%%%
In this framework, the extra spatial dimension, represented by the radius $R_y$, is compactified on the $S_1/Z_2$ orbifold. The warped solution of the metric arises due to the nonzero bulk cosmological constant $\Lambda$, which satisfies, $k \equiv \sqrt{-\frac{\Lambda}{24 M^3}}$. Consequently, the bulk geometry corresponds to a 5 dimensional anti-deSitter spacetime. Here $k, \,M$ denote the bulk curvature and fundamental Planck mass in five dimensions.    
Two 3-branes are located at the orbifold fixed points such as $y=0,\,\pi$, separated along the extra dimension and known as the Planck (hidden) and TeV (visible) brane respectively. 
The exponential warp factor associated with the fifth dimension generates a natural hierarchy between the Planck and electroweak scales by exponentially suppressing mass parameters on one brane relative to the other. Consequently, any mass parameter $m_0$ defined on the $(3+1)$ dimensional visible brane in the underlying higher dimensional theory corresponds to a physical mass given by $m=m_0\,e^{-k R_y\,\pi}$. This exponential suppression provides a natural resolution to the Higgs mass hierarchy problem, which arises from the sensitivity of the Higgs mass to the radiative quantum corrections.
%%%
%%%
%%

In the original RS model, the SM fields are assumed to lie on the TeV brane while only gravity propagates in the bulk. Subsequent works have investigated the phenomenology of SM fields propagating in the bulk of warped geometry models \cite{Davoudiasl:1999tf,Huber:2000fh}. In the present work, we consider a massless scalar field propagating in the bulk and derive the effective scalar field action along with the KK mass spectrum on the ($3+1$) dimensional visible brane.
Although our primary interest is in a massless scalar field, we begin with the more general case of a 5 dimensional massive scalar field of mass $m_0$, with the action given by, 
%%%
{\small
\beq
% \adjustbox{max width=\columnwidth}{$\displaystyle
S_\Phi=\,\frac{1}{2}\int d^4x \int_{-\pi}^\pi dy\,\sqrt{-g^{(5)}}(g^{MN}\partial_M\Phi\,\partial_N\Phi-m_{0}^{2}\,\Phi^2)
% $}
\label{scalar_action_rs_1}
\eeq
In the above action we use the KK mode decomposition for the bulk scalar field as follows,
}
%%%%
\begin{equation}
    \Phi(x^{\mu},y)=\sum_n\frac{1}{\sqrt{R_y}}\phi_n(x^\mu)\xi(y)\,
    \label{scalar_kk_rs}
\end{equation}
and integrating with respect to the $y$ coordinate, one obtains the effective action as,
%%%%%
\begin{equation}
S_{(4)}=\,\frac{1}{2}\sum_n\int d^4x\,\left(\eta^{\mu\nu}\,\partial_{\mu}\phi_n\,\partial_{\nu}\phi_n-\,m_{n}^{2}\phi_{n}^{2}\right)\label{scalar_action_rs_2}\,,
\end{equation}
%%%%
To obtain Eq. (\ref{scalar_action_rs_2}), we employ the normalization condition as, 
\beq
\int_{-\pi}^{\pi}\,e^{-2 k R_{y}\,|y|}\,\xi_{m}(y)\,\xi_{n}(y)\, dy = N_n^2\,\delta_{mn} 
\label{norm_rs_1}
\eeq
and the differential equation corresponding to the $y$ coordinate becomes, 
\begin{align}
-\f{1}{R_{y}^{2}}\f{d}{dy}\left[e^{-4 k R_{y}|y|}\f{d\xi_n(y)}{dy}\right]+m_{0}^{2} e^{-4 k R_{y} |y|}=m_{n}^{2}e^{-2 k R_{y}|y|}\xi_n(y)\,.
\label{dynamics_y}
\end{align}
%%%%
Here $m_n=x_{n}k\,e^{-kR_y\pi}$ depict the KK mass modes for the scalar field as induced on the visible brane. $x_{n\nu}$ satisfies $2J_{\nu}(x_{n\nu})+x_{n\nu}J_\nu^\prime(x_{n\nu})=0$, where $J_\nu$ is the Bessel function of the order $\nu = \sqrt{4+m_{0}^{2}/k^2}$. $N_n$ is the normalization constant,
\begin{equation}
    N_n = \frac{e^{kR_y\pi}}{\sqrt{kR_y}}\left(\sqrt{1+\frac{4-\nu^2}{x_{n\nu}}}\right)J_\nu(x_{n\nu})
\end{equation}
%%%%
Following the procedure as described in the Secs. (\ref{res_ud_1}), (\ref{kms}), (\ref{spont_massive}) and (\ref{flat_spectrum_1}), we obtain the profile function of the detector for the warped geometry model as follows,
%
% \begin{align}
% \mathcal{F}(\D E)&=\,\D E+\sum_{n=1}^{\infty}\left[ \sqrt{(\D E)^2-x_{n}^2\,k^2\,e^{-2 k R_y \pi}}\notag\\
% &\Theta\left(\D E-x_{n}\,k\,e^{-k R_y \pi}\right) \right]\,.
% \label{RS response function}
% \end{align}
\begin{align}
\mathcal{F}(\Delta E) &= \Delta E + \sum_{n=1}^{\infty} \Biggl[ \sqrt{(\Delta E)^2 - x_n^2\, k^2\, e^{-2kR_y\pi}} \notag \\
&\hspace{2.2cm} \times\, \Theta\left(\Delta E - x_n\, k\, e^{-kR_y\pi}\right) \Biggr] \, .
\label{RS response function}
\end{align}
%%%%%
The emission and absorption rate becomes, 
\begin{align}
\dot{F}(\D E)=\f{1}{e^{2 \pi \D E/a}-1}&\Biggl[\D E+\sum_{n=1}^{\infty}\sqrt{(\D E)^2-x_{n\nu}^2 k^2 e^{-2 k R_y \pi}}\notag\\
&\times \Theta\left(\D E-x_{n\nu} \,k\,e^{-k R_y \pi}\right)\Biggr]\,.
\end{align}
% \begin{equation}
%    \dot{F}(\D E) = \frac{\D E+ \sum_{n=1}^{\infty}\left[ \sqrt{(\D E)^2-x_{n\nu}^2\,k^2\,e^{-2 k R_y \pi}}\,\,\Theta\left(\D E-x_{n\nu} \,k\,e^{-k R_y \pi}\right)\right]}{e^{2 \pi \D E/a}-1}
% \end{equation}
%%%
Eq. (\ref{RS response function}) depicts the master equation to examine the profile function of the UD detector in the background of RS warped geometry model. 

\noindent
Generalizing 5 dimensional RS scenario to 6 dimensional doubly warped geometry model, we compute the profile function for the UD detector as follows, 
%%%%
% \begin{align}
%     \mathcal{F}(\D E)= & \mathlarger{\sum}_{n,q=0}^{\infty}\sqrt{(\D E)^2- \left(\frac{x_{\nu _n}k\alpha}{R_y}\right)^2\frac{1}{\cosh^2(k\pi)} {\rm exp}\left[\frac{- 2k\alpha\pi}{\cosh(k\pi)}\right]}\notag\\
%     &\times\Theta\left[\D E- \frac{x_{\nu _n} k\,\alpha}{R_y\,\cosh(k\pi)}\,{\rm exp}\left(\frac{- k\,\alpha\,\pi}{\cosh(k\pi)}\right)\right]\,.
%     \label{profile_fun_rs_6}
% \end{align}
%%%%
\begin{align}
    \mathcal{F}(\D E)= & \mathlarger{\sum}_{n,q=0}^{\infty}\sqrt{(\D E)^2- \left(\frac{x_{\nu _n}k\alpha}{R_y}\right)^2\frac{e^{-2 k\a \pi/\cosh(k\pi)}}{\cosh^2(k\pi)}}\notag\\
    &\times\Theta\left[\D E- \frac{x_{\nu _n} k\,\alpha}{R_y\,\cosh(k\pi)}\,{\rm exp}\left(\frac{- k\,\alpha\,\pi}{\cosh(k\pi)}\right)\right]\,.
    \label{profile_fun_rs_6}
\end{align}
In the above equation $\a=R_y/r_z$, where  $R_y,\,r_z$ denote the radii of the extra dimensions along $y,\,z$ directions respectively. $k$ represents the characteristics warping along the $z$ direction and $x_{\nu_{n}}$ are the roots of the Bessel function as depicted in Eq. (\ref{eq:bessel_6d_rs_1}). For the detailed derivation of the profile function, in case of the 6D RS model, we refer readers to the Appendix (\ref{rs_appendix_1}). 
%%%%
\FloatBarrier
\section{Parametric variation of the profile function in the large and flat extra dimensional scenario}\label{plot_add}
%%%
We study the profile function of the detector in the background of the large and flat extra dimensional model such as ADD framework. However, it is worth mentioning couple of points as below: 
\begin{itemize}
    \item 
From $R\sim 10^{32/d} \,{\rm TeV}^{-1}$, ADD model with one extra dimension predicts the radius of the extra dimension to be $\mathcal{O}\left(10^{32}\right)\,{\rm TeV}^{-1}\sim 10^{15}\,{\rm cm}$, if $M_{*}$ is restricted to be $\sim \mathcal{O}(1)\,{\rm TeV}$ in order to resolve the hierarchy problem.  
%%%%%
%%%%%
This radius value exceeds the separation between the earth and the sun ($\sim 10^{13}\,{\rm cm}$), which implies that there should be an observable deviations from the Newtonian $1/r^2$ law within the solar system. Since precision tests of gravity such as from the planetary dynamics to submillimeter torsion balance experiment show no such deviation, one concludes that the ADD model with one extra dimension is ruled out. 
\item
Despite being observationally ruled out, we consider $d=1$ case and its consequences for illustrative purposes in our manuscript. $d=1$ case provides a simplest analytical setting to examine the behaviour of the profile function with various parameters of the system. 
\item
To generate the plots, for $d=1$, we consider $R \ll 10^{32}\,\,{\rm TeV}^{-1}$ which jeopardizes the resolution to the hierarchy problem, as $M_{*}$
no longer coincides with the electroweak scale. Nevertheless, this restricted setting remains valuable for studying the behaviour of the profile function of the UD detector, particle production phenomena, Unruh type thermal effects, detector's sensitivity to the KK mass modes and size of the extra dimension etc. These aspects are largely independent of whether the hierarchy problem is fully resolved. 
Thus, $d=1$ case offers a simple framework to understand these interactions before extending to the phenomenologically viable scenario {\it i.e.}, $d\geq 2$. 
\item 
For $d=2$, $R \sim 10^{16}\,\,{\rm TeV}^{-1}$ to solve the hierarchy problem. Although, setting other values for $R$ affect the resolution of the hierarchy problem, our primary aim remains as to observe the characteristic behaviour of the profile function, dependence of the same on the dimensionality of the spacetime. Therefore we set $R\sim \mathcal{O}\left(10^{12}\right)\,\,{\rm TeV}^{-1}$ and plot Eq. (\ref{profile_function_add_1}) for $d=2$ dimensional ADD scenario. 
\end{itemize}
Henceforth we use $\D E =E$ for the brevity of notations. 
%%%%
\subsection*{$\mathcal{F}(E)$ vs. $E$ and $\mathcal{F}(E)$ vs. $R$ plots for $d=1$ in the large and flat extra dimension}
%%%%%%%%%%%%%%
\begin{figure}[htbp]
    \centering    \includegraphics[width=0.72\columnwidth]{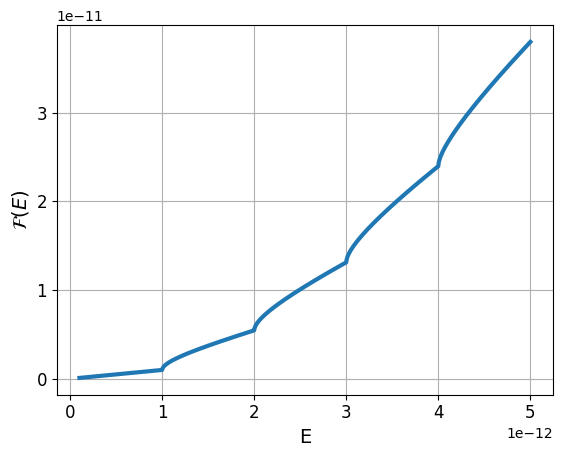}
    \vspace{0.3cm}
    \includegraphics[width=0.72\columnwidth]{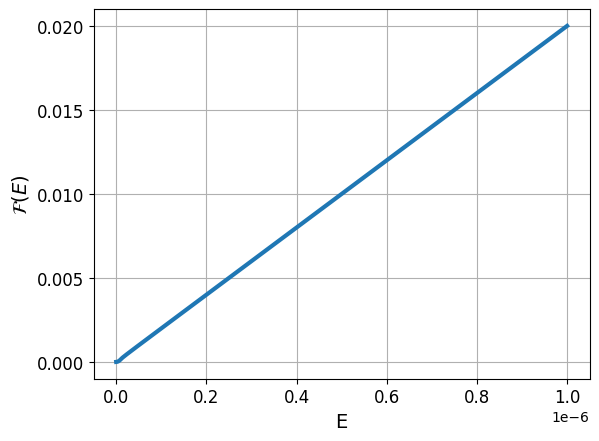}
    \caption{For $d=1$ and $R=10^{12}\,\mathrm{TeV}^{-1}$, we plot $\mathcal{F}(E)$ as a function of $E$ for the flat and large extra-dimensional scenarios. In the top and bottom panel, $E$ varies over the ranges $0$--$5\times10^{-12}\,\mathrm{TeV}$, $0$--$1\times10^{-6}\,\mathrm{TeV}$, respectively.}
    \label{fig:ml_add_e_1}
\end{figure}
%%%%%%%%%%
%%%%%%%%%%%%%%%
\begin{figure}[htbp]
    \centering
\includegraphics[scale=0.44]{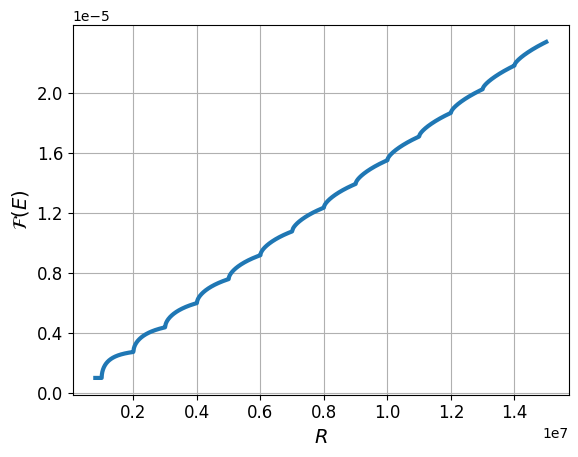}
\vspace{0.3cm}
        \includegraphics[scale=0.44]{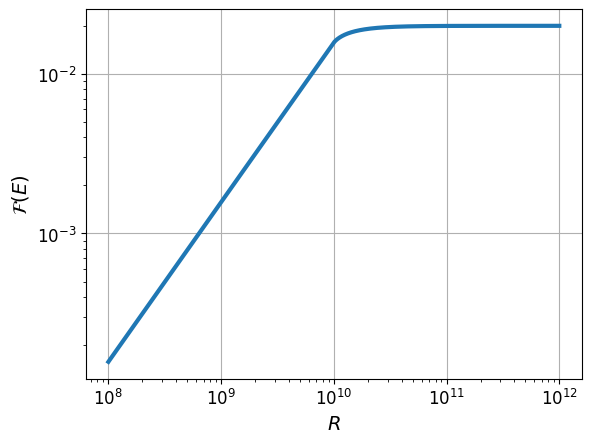}
 \caption{Top panel and bottom panel show the profile function as a function of $R$ for $E=1\,\mathrm{MeV}$ and $d=1$. The $(R,\,E)$ axes are plotted with the unit ${\rm TeV}^{-1},\,{\rm TeV}$ respectively. }
\label{fig:flat_combined}
\end{figure}
%%%%%%%
%We summarize the Figs. (\ref{fig:ml_add_e_1}) and (\ref{fig:flat_combined}) as below: 
\begin{enumerate}
   \item Eq.~(\ref{kk_pro_fun_1}) shows that a given KK mass mode $m_n$ can only be excited once the detector's energy gap satisfies $E \geq m_n$, imposed by the step function $\Theta(E - m_n)$. Consequently, as $E$ is increased, a larger number of KK modes become accessible,  so comparing top and bottom panel of Fig. (\ref{fig:ml_add_e_1}), we obtain a profile function that grows with $E$.
    \item 
    The top panel of Fig. (\ref{fig:ml_add_e_1}) exhibits step like features (kinks), whereas the bottom panel shows comparatively smoother pattern. This follows directly from Eq. (\ref{kk_pro_fun_1}). The energy range in the top panel is a few multiples of the KK mass spacing $1/R$. Hence successive KK modes,  $m_n=n/R$ are individually detected as $E$ crosses each one. In the bottom panel of the same, energy range spans a much larger number of KK mass spacings, so a large number of modes turn on with rapid detection. This leads to the respective rough and smooth pattern of these two plots. 
%%%%%
\item In the top panel of Fig. (\ref{fig:flat_combined}), for $R\gtrsim 10^{6}\,{\rm TeV}^{-1}$, the profile function shows staircase like growth due to the successive detection of KK mass modes. However, as  $R$ increases, the profile function becomes comparatively smoother. This behavior arises from the variation in the mass gap between successive KK modes. 
Smaller values of $R$, corresponding to larger mass gaps, lead to a more irregular (staircase like) profile, whereas larger $R$, corresponding to smaller mass gaps, result in a smoother profile function.
\item 
In the bottom panel of Fig. (\ref{fig:flat_combined}), we observe an apparent saturation for higher $R$ values, with no further detection of new KK modes. This behaviour arises from the computational limitation in evaluating the summation over KK modes as $n\to\infty$. Thus for the sake of computational simplicity, we truncate the summation at $n\sim\mathcal{O}(10^3-10^4)$. Hence, the observed saturation should not be interpreted as the actual asymptotic behaviour of the profile function, which would continue to evolve if the summation could be extended to $n\to\infty$.
\end{enumerate}
%%%%%
\subsection*{$\mathcal{F}(E)$ vs. $E$ and $\mathcal{F}(E)$ vs. $R$ plots for $d=2$ in the large and flat extra dimension} 
%%%%
\begin{figure}[htbp]
        \centering    \includegraphics[width=0.72\columnwidth]{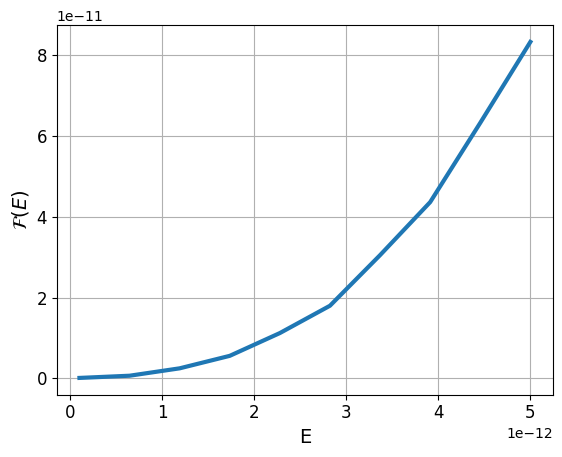}
     \vspace{0.3cm}
\centering
\includegraphics[width=0.72\columnwidth]{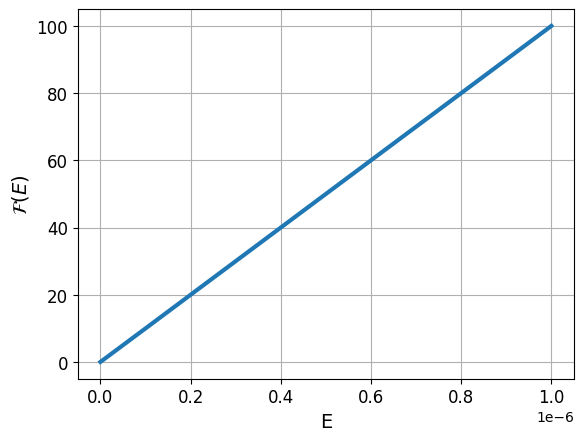}     
    \caption{$\mathcal{F}(E)$ vs $E$ plots for two flat extra dimensions while $R=10^{12}$ TeV$^{-1}$.}
    \label{2d_variation_E}
    \end{figure}
%%%%%%%%%%%%%%%%%%%%%%%%%%%%%
\begin{figure}[htbp]
    \centering
    %%% r plot 1
\includegraphics[width=0.74\columnwidth]{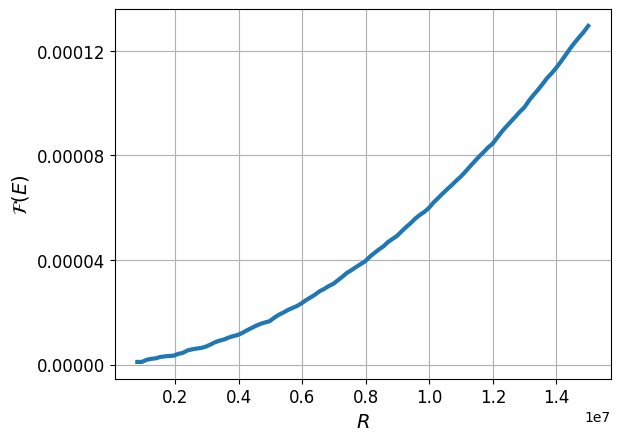}
     \vspace{0.3cm}
     \centering
    \includegraphics[width=0.72\columnwidth]{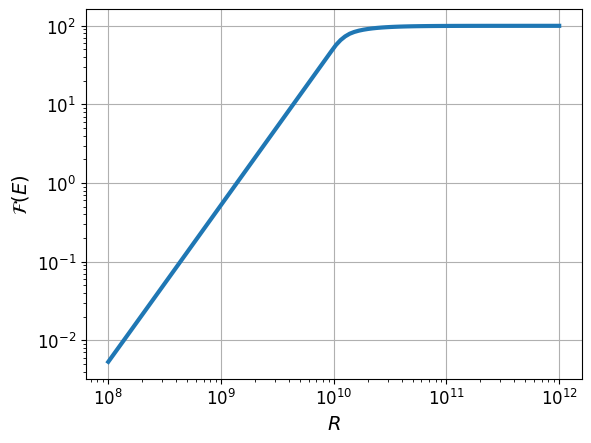}
    \caption{$\mathcal{F}(E)$ vs $R$ plots for two flat extra dimensions while fixing the energy gap at $E= 10^{-6}$ TeV.} 
    \label{fig:2_flat_1}
\end{figure}
%%%%
%Summary of Figs. (\ref{2d_variation_E}) and (\ref{fig:2_flat_1}):
\begin{enumerate}
\item
The top and the bottom panels of Figs. (\ref{fig:ml_add_e_1}) and (\ref{2d_variation_E}) show an increase in the profile function with increasing $E$. However, comparing both the figures it can be seen that the growth of the profile function is smoother in the latter. This follows from Eq. (\ref{profile_function_add_1}), where for two compact extra dimensions, the KK modes are labelled by a pair of integers $(n,q)$ rather than a single integer $n$. Consequently, more combinations of KK modes satisfying $m_{nq}<E$ become accessible within the same plotted energy range. This leads to a higher density of accessible modes and therefore the top panel of Fig. (\ref{2d_variation_E}) appears smoother than that of the same panel of Fig. (\ref{fig:ml_add_e_1}). 
 A similar comment applies to the bottom panel plots of the two figs. (\ref{fig:ml_add_e_1}) and (\ref{2d_variation_E}).
\item 
Comparing the top panels of the Figs. (\ref{fig:flat_combined})
and (\ref{fig:2_flat_1}) we obtain an enhanced and smoother profile function for the $d=2$ case. The bottom panel of these figures depicts a similar enhanced plot for the profile function in 6 dimensions. The bottom panel plots become more smoother than the corresponding top panel plots of these figures due to the choice of the higher values of the radii of the extra dimensions. 
 \end{enumerate}
 %%%
 %%
 \section{Parametric variation of the profile function in the warped extra dimensional scenario}\label{plot_rs}
%%%%
We examine the behaviour of the profile function in the background of RS warped geometry model and highlight the following points:
\begin{figure}[htbp]
    \centering
\includegraphics[width=0.72\columnwidth]{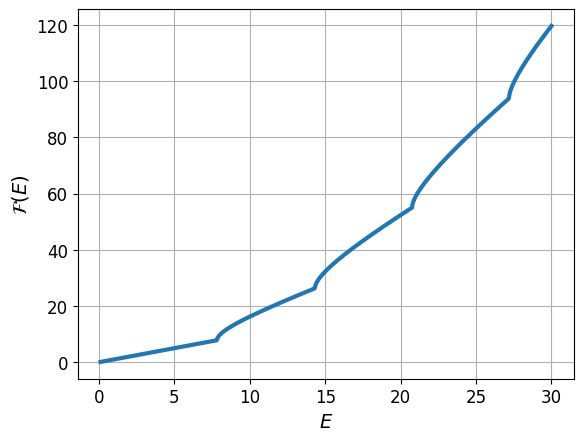}
      \vspace{0.3cm}
      \centering
   \includegraphics[width=0.72\columnwidth]{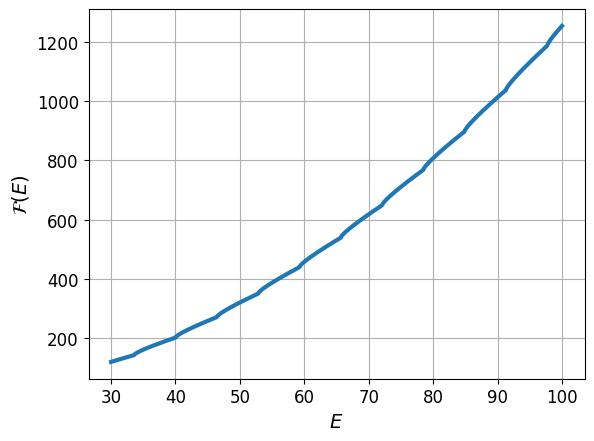 }
     \caption{Profile function vs E for a single warped extra-dimensional model with $R_y=10^{-15}$ TeV\(^{-1}\). $E$ and $\mathcal{F}(E)$ are in TeV unit. } 
    \label{fig:ml_rs_e_1}
\end{figure}
%%%%%%%%%%%%%%%%%%%%%
%%%%%%%%%%%%%%%%%%%%%
\begin{figure}[htbp]
    \centering
    \includegraphics[width=0.73\columnwidth]{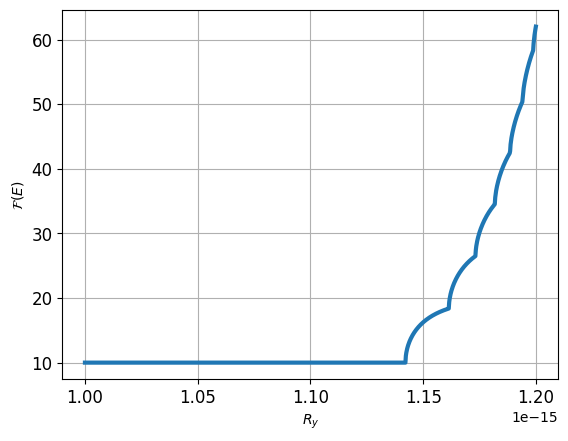}
    \vspace{0.3cm}
    \centering
    \includegraphics[width=0.73\columnwidth]{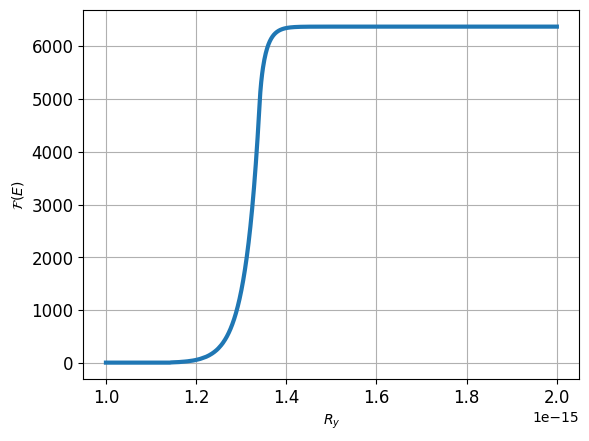}
    \centering
\vspace{0.3cm}
\centering
    \includegraphics[width=0.73\columnwidth]{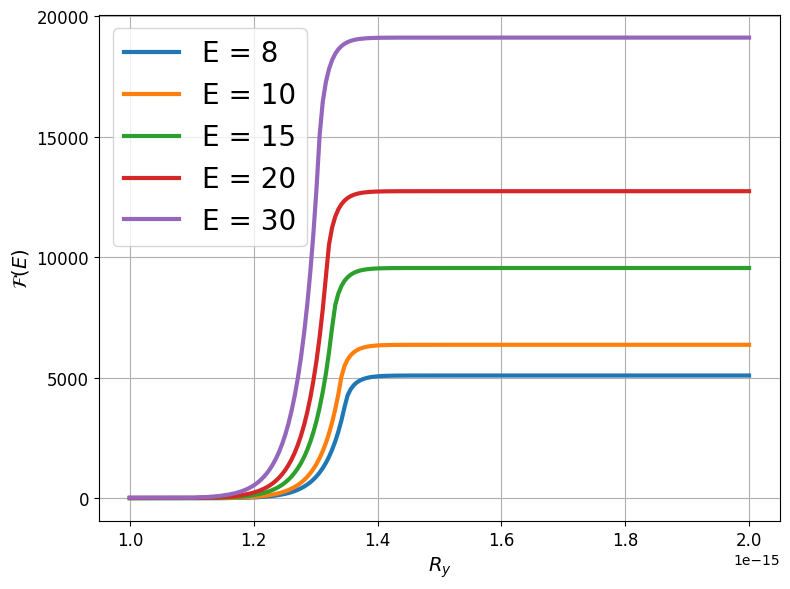}
    \caption{Top panel: $\mathcal{F}(E)$ vs $R_y$ plots for $d=1$ warped geometry model where $E = 10$ TeV. Bottom panel: $\mathcal{F}(E)$ vs $R_y$ plots for different values of $E$.}
    \label{fig:ml_rs_r_combined}
\end{figure}
%%%%%
\begin{enumerate}
\item In case of RS scenario, the resolution of the hierarchy problem suggests the value of $k R_y \sim 12$. This leads us to fix the $R_y \sim \mathcal{O}\left(10^{-15}\right)\,\,{\rm TeV}^{-1}$. Furthermore, this value of the extra dimensional modulus is also consistent with the Goldberger-Wise stabilisation mechanism \cite{Goldberger:1999wh}. Therefore, fixing $R_y\sim \mathcal{O}\left(10^{-15}\right)\,{\rm TeV}^{-1}$, which addresses the hierarchy problem, we plot the profile function as a function of $E$ in Fig.~(\ref{fig:ml_rs_e_1}). 
The top and bottom panel of the Fig.~(\ref{fig:ml_rs_e_1}) show that the profile function increases with the energy gap of the detector, as a larger energy gap enables the detection of a greater number of KK mass modes. 
    \item The staircase-like structure in the top and bottom panel of Fig.~(\ref{fig:ml_rs_e_1}) does not smooth into a continuum at higher $E$. 
%For the flat and large extra dimensional model, the KK mass spacing $1/R$ is fixed and parametrically tiny due to large $R$. Thus for sufficiently large $E$, an enormous number of KK modes lie below it, producing an effectively continuous profile function. 
In the RS scenario, the roots of the Bessel equation $2J_{\nu}(x_{n\nu})+x_{n\nu}J_\nu^\prime(x_{n\nu})=0$ {\it i.e} $x_n$ and $m_n = x_n k\, e^{-kR_y\pi}$, become asymptotically equally spaced for large $n$. Therefore, the KK mass spacing approaches a constant value, rather than decreasing as $n$ increases. This gives $\D m\approx 6.4-6.5$ TeV, in agreement with the spacing between $m_1$ to $m_4$ in Table~(\ref{tab:5wed mass modes}). As a result, the discrete structure of $\mathcal{F}(E)$ persists across the plotted range in Fig.~(\ref{fig:ml_rs_e_1}). We comment that this discrete pattern of $\mathcal{F}(E)$ in both the models, depend on the range over which the energy gap of the detector is sampled. 
    %%%%%
\item In Fig.~(\ref{fig:ml_rs_r_combined}), we plot $\mathcal{F}(E)$ as a function of $R_y$. The profile function remains flat for $R_y\sim (1-1.15)\times10^{-15}\,\mathrm{TeV}^{-1}$, indicating no detection of KK modes in this range and increases sharply thereafter. 
\item 
In the flat model, the threshold radius for the $n$-th KK mode varies linearly with $n$, which for the threshold condition of the mass detection satisfy $E=m_n=n/R$ and yields $R=n/E$. Therefore, successive thresholds are spread over a range of $R$ that grows in direct proportion to the number of modes probed. Consequently, the profile function rises gradually in a staircase pattern. In contrast for RS model, the threshold radius scales only logarithmically with $x_n$, $R_{y}\propto\ln (x_n)$ due to the exponential warp factor $e^{-kR_y\pi}$. Therefore, an exponentially large number of KK modes cross threshold within a narrow window of $R_y$, producing the sharp, collective rise as seen in the top bottom panel of Fig.~(\ref{fig:ml_rs_r_combined}).
\\
\noindent
This difference is a direct consequence of the underlying geometry rather than a numerical coincidence and it allows the UD detector to distinguish between the two classes of extra dimensional scenarios, purely based on the profile function of the detector without requiring prior knowledge of the compactification scale.  
%%%%
\end{enumerate}
%%%%
\subsection*{$\mathcal{F}(E)$ vs. $E$ and $\mathcal{F}(E)$ vs. $R_y$ plots for $d=2$ warped extra dimensional model}
%%%
We plot Eq. (\ref{profile_fun_rs_6}) with $E$ and $R_y$ while keeping $k$ and $\a$ fixed. Setting $\a \sim 
\mathcal{O}(60)$, confirms that $R_y,\,r_z$ are not hierarchical in their values. 
%%%%%%
\begin{figure}[htbp]
    \centering
\includegraphics[scale=0.43]{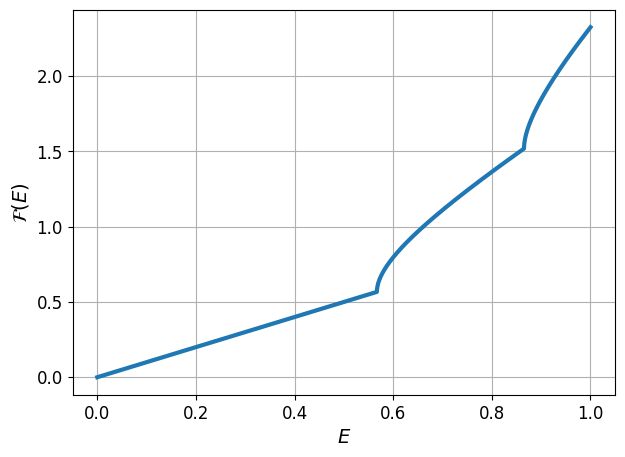}
\vspace{0.3cm}
\centering
\includegraphics[scale=0.43]{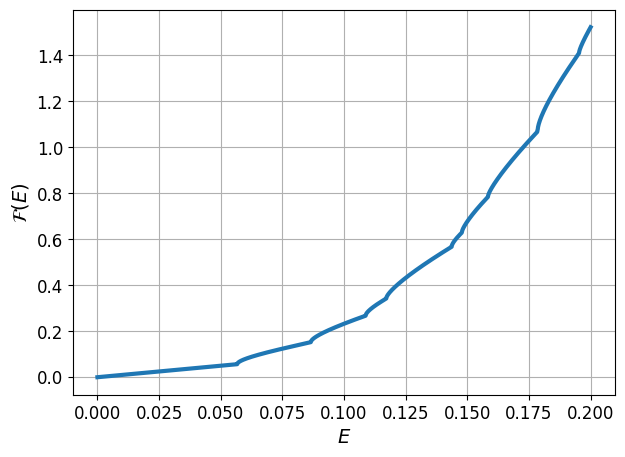}
    \caption{$\mathcal{F}(E)$ vs $E$ plots for $d=2$. In the top panel, we have kept $R_y=10^{-14}\,{\rm TeV}^{-1}$ and in the bottom panel $R_y=10^{-13}\,{\rm TeV}^{-1}$. For both the figures $k=0.25$ TeV, $\alpha = 61.6605$ and $M/k'=0$.}
    \label{fig:ml_6wed_e}
\end{figure}
%%%%%
\begin{enumerate}
\item 
Comparing Figs. (\ref{2d_variation_E}) and (\ref{fig:ml_6wed_e}), we observe a similar feature that increasing the number of spacetime dimensions leads to a denser accumulation of KK modes, due to the enhanced splitting of their masses. Consequently, a larger number of KK modes become accessible to the UD detector which in turn enhances its profile function. 
%%%%%
We also observe that increasing the number of spacetime dimensions reduces the energy gap required for the UD detector to achieve a given profile function. This reduction in the required energy gap makes the detection of KK modes more experimentally accessible.
%%%%
\begin{figure}[htbp]
\centering 
\includegraphics[scale=0.4]{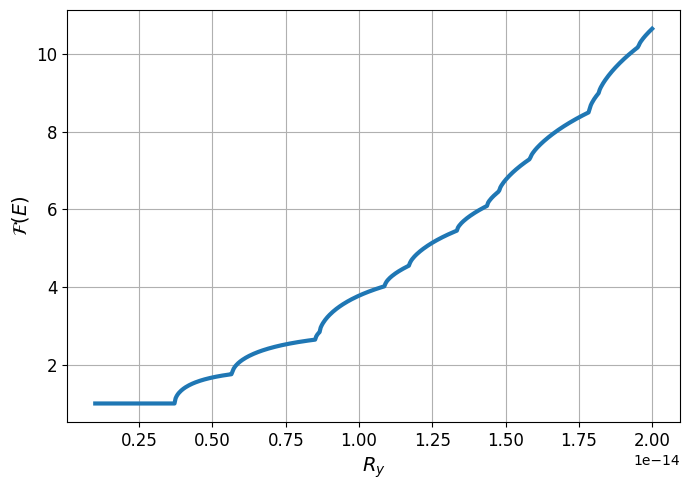}
\vspace{0.3cm}
\centering
    \includegraphics[scale=0.4]{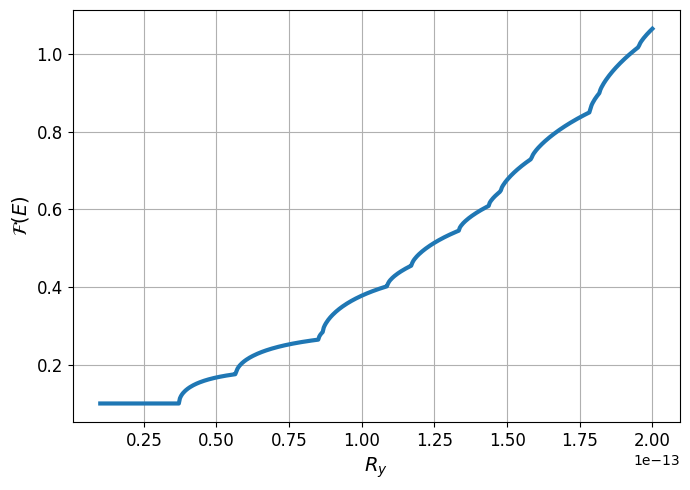}
\caption{Profile function vs $R_y$ for two warped extra dimensions keeping  $k=0.25$ TeV, $\alpha = 61.6605$. In the top panel, $E=1$ TeV, and in the bottom panel $E = 0.1$ TeV.}
    \label{fig:ml_6wed_r_all}
\end{figure}
\item
Comparing both the plots in Fig. (\ref{fig:ml_6wed_e}), it can be noted that increasing the magnitude of the extra dimensional radii for $d=2$ case, leads to the reduction in the KK mass modes. Therefore in order to achieve a particular $\mathcal{F}(E)$, one obtains a reduction in the required energy gap of the detector for the larger radii case (Ex: bottom panel).  A similar pattern can be observed in Fig. (\ref{fig:ml_6wed_r_all}) where obtaining a profile function that is not overly suppressed as it varies with $R_y$, requires two different values of $E$ to be set. 
%%%%%
\end{enumerate}
%%%%%%%%%%%%
%%%%%%%%%%%%%
\subsection{Comparison of standard and extra dimensional profile functions}
We compare the behaviour of the profile function with the energy gap of the detector for the standard UD detector and extra dimensional models. 
%%%%
\begin{figure}[htb!]
    % \centering
    %    \includegraphics[scale=0.37]{corrected_plots_final/union_1.png}
    %    \vspace{0.3cm}
\centering
\includegraphics[scale=0.37]{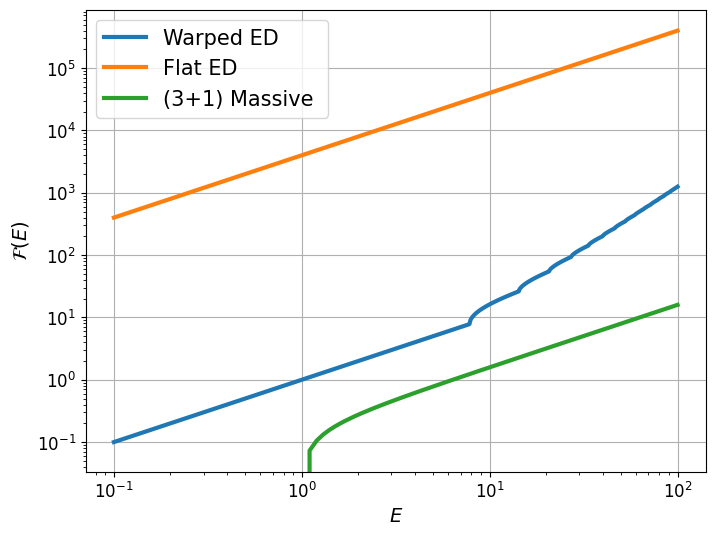}
% \vspace{0.3cm}
% \centering
% \includegraphics[scale=0.37]{corrected_plots_final/union_3.png}
    \caption{Comparison of the profile function as a function of energy gap for (single) extra dimensional models and $3+1$ dimensions. For flat extra dimension model, $R=10^12$ TeV$^{-1}$. For warped extra dimension model, $k=10^{16}$ TeV and $kR_y=11.5$. For the (3+1) scalar field model, $m=1$ TeV. The axes of all the plots are in logarithmic scales with base 10. }
    \label{fig: Union profile functions}
\end{figure}
%%%%%
We observe a larger amplitude in the profile function for the large and flat extra dimensional model in comparison to warped extra dimensional and standard UD detector models. For the standard case we consider $1$ TeV as the mass of the scalar field which shows the detection of the field quanta around $1$ TeV. 
%%%%%%%
\section{Discussion}\label{discussion}
%\subsection*{some important fundamental lines}
In this work, we compute and examine the profile function $\mathcal{F}(E)$ of UD detector for two distinct classes of extra dimensional models, such as flat and large extra dimensional models (Ex: ADD model) and warped extra dimensional scenario (RS model). Furthermore, we compute the same for the 6 dimensional generalizations of these models and examine their behaviour with respect to the energy gap of the detector, radius of the extra dimension, and number of spacetime dimensions. Throughout the manuscript, we plot the profile function $\mathcal{F}(E)$, which in turn is related to the response rate ($\dot{F}(E)$) of the detector as depicted in Eq. (\ref{abs_final_1}). The thermal factor in Eq. (\ref{abs_final_1}), is smooth and monotonic in $E$, and it is not affected by the compactification geometry of the extra dimension. Thus, all the characteristics such as kinks and staircase like patterns, captured in $\mathcal{F}(E)$, remain intact in the physically measurable quantity $\dot{F}(E)$. The thermal factor in $\dot{F}(E)$ only regulates the overall magnitude. 
%%%%%
The distinctive features in the profile function arise due to the difference in the resolution of the threshold/ kink points where $E=m_n$. For flat/ ADD model, the gap between successive threshold is $\Delta m_n \sim 1/R$ and at the same time the resolution of hierarchy problem requires $R \sim 10^{32/d}\,{\rm TeV}^{-1}$, resulting $1/R$ to be very small parameter. However, kinks are distinguishable based on the restricted resolution as depicted in the top panel of figs. (\ref{fig:ml_add_e_1}) and (\ref{fig:flat_combined}). As $E$ increases, one accommodates more and more threshold points, as the spacing of the KK mass modes remain fixed. Thus, beyond a certain energy scale, 
successive kinks get densely spaced so that they cannot be resolved individually. This yields a smooth curve for the profile function of the detector with increasing $E$ [see bottom panels of figs. (\ref{fig:ml_add_e_1}) and (\ref{fig:flat_combined})]. For six dimensional flat/ ADD scenario we obtain smoother and enhanced plots in the profile function for the same parametric variations. This implies the detection of more KK modes due to the mass splitting in the higher dimensions, as discussed in Sec. (\ref{plot_add}). 
%%%%%
In case of RS scenario the gap in the KK mass modes for large $n$ approaches a constant value due to the Bessel equation $2J_{\nu}(x_{n\nu})+x_{n\nu}J_\nu^\prime(x_{n\nu})=0$, as described in Sec. (\ref{plot_rs}). Therefore the discrete pattern remains identifiable even for large energy gap of the detector [see Fig. (\ref{fig:ml_rs_e_1}) and the top panel of Fig. (\ref{fig:ml_rs_r_combined})]. However, this pattern is subject to the choice of the energy resolution of the detector {\it i.e}, the range over which the energy gap is scanned.
%%%%
In case of the radius of the extra dimension, for the flat/ADD model, the threshold radius $R$ varies linearly with KK modes $n$ ({\it i.e}., $R=n/E$), whereas, $R_y$ varies as $\sim {\rm ln}\,x_n$ in RS warped geometry model. This results a gradual increment in $\mathcal{F}(E)$ while varying with respect to the $R$ for the large and flat extra dimensional model. On contrary, a tiny variation in $R_y$, simultaneously accommodates large number of RS KK modes due to the exponential characteristics of the warp factor, yielding a sharp increase in $\mathcal{F}(E)$ [see Fig. (\ref{fig:ml_rs_r_combined})]. 
%%%%%
\\
\noindent
Our parametric analysis further demonstrates that, for a given value of $\mathcal{F}(E)$, the required detector energy gap ($E$) decreases with increasing number of extra dimensions, in both the ADD and RS warped geometry models. This feature arises from the larger number of KK mass modes that become accessible at a given energy $E$, as additional compactified dimensions are introduced.
%Thus we comment that models with a larger number of compact dimensions require a relatively small energy gap of the detector to probe their characteristic KK signature.
This reduction in the required energy scale may improve the experimental accessibility of the corresponding KK signatures. In addition, the KK tower produces distinctive modifications of the detector response, including an enhanced magnitude of the response function and geometry dependent threshold structures. Since the profile function is directly related to the physically measurable response rate \(\dot F(E)\), these features can be investigated through a controlled variation of the detector energy gap. Notably, the contrasting pattern of the response function as obtained for flat and warped extra dimensions suggest that the detector response may provide an indirect signature of the underlying compactification geometry.
%%%%%
\\
\noindent
To the best of our knowledge, within the detector based formalism for extra dimensional models, as reported in the literature, the present work for the first time provides a thorough derivation and analysis of the profile function of the UD detector in a warped extra dimensional geometry, specifically the RS scenario. Furthermore, our work presents a detailed analysis of models with multiple extra dimensions, considering both large, flat and warped geometry models. For both models, we systematically examine the profile function of the detector by varying the compactification radius of the extra dimension, energy gap of the detector, and number of extra dimensions, while also investigating the implications for the resolution of the hierarchy problem. Consequently, the parameter space considered in our analysis is not arbitrary, rather, it is guided by whether the corresponding parameter choices allow for the resolution of the hierarchy problem.
%%%%%
Fig. (\ref{fig: Union profile functions}) compares the flat, warped extra and standard $(3+1)$ dimensional profile function for UD detector interacting with massive scalar field. 
This comparison projects the primary distinguishing feature between the two extra dimensional classes. The flat scenario produces a larger profile function, as it gathers a larger number of closely spaced KK modes within the chosen range of the energy gap. On contrary, RS scenario shows a sharper and discrete response due to its irregularly spaced KK modes. 
%%%%%%
These results extend the UD detector framework beyond the single, standard flat spacetime scenario, as studied  previously in the literature. 
The present study shows that the UD detector formalism can provide a useful tool for distinguishing between flat and warped extra dimensional geometries while using the values of the parameters as consistent with the resolution of the hierarchy problem.
Therefore we comment that the UD detector can capture the signature of the underlying extra dimensional geometries within its profile function.     
Our results, as presented in this manuscripts, can be considered as an ideal probe of extra dimensional geometry, offering a theoretically consistent way to isolate the effect of compactification structure of the  two promising extra dimensional models, with respect to the response of the UD detector.

\appendix
%\FloatBarrier
\section{Derivation of the response rate of the UD detector}\label{appendix_2}
From eqs. (\ref{eq:res_fun_gen_1}) and (\ref{eq: Wightman function}), the response function and Wightman function are respectively
% The transition amplitude and probability of going from an initial state $(E_i,0_M)$ to a final state $(E_f,1_p)$ using the first order perturbation theory is 
% \begin{align}
%     A_{i\to f}&= -\int d\tau \braket{E_f,1_p|g\,\mu(\tau)\phi(x(\tau))|E_i,0_M}\,,
%   \label{eq: Transition amplitude}
%   \end{align}
%   %%%%%
% which leads to the transition probability as follows, 
%   \begin{align}
%       P_{i\to f}= \int d^3p\,|A_{i\to f}|^2 = CF(\Delta E)\,,
%   \end{align}
% where $C= g^2|\braket{E_f|\mu(0)|E_i}|^2$, $\D E=E_f-E_i$. 
\begin{align}
F(\Delta E) &= \int d\tau d\tau^\prime e^{-i\Delta E(\tau-\tau')}\braket{0_M|\phi(x')\phi(x)|0_M}\notag\\
&=\, \int d\tau d\tau^\prime e^{-i\Delta E(\tau-\tau')}G_+(\tau,\tau')\label{re_fn_gen_1}
\end{align}
%%%
and
%%%
% In the above expression, $G_+(x,x')=\braket{0_M|\phi(x')\phi(x)|0_M}$ represents the Wightman function.
% given by
\begin{align}
&G_+= \braket{0_M|\phi(x')\phi(x)|0_M}\\
    &=\int \frac{d^3\bar{p}}{(2\pi)^3}\frac{e^{i(\bar{p}.(x(\tau')-x(\tau))-\omega_p (t(\tau')-t(\tau))}}{2\omega_p} \notag\\
    &=\int \frac{d^3\bar{p}}{(2\pi)^3}\frac{e^{i(\bar{p}.\Delta x-\omega_p \Delta t)}}{2\omega_p}\label{eq: Wightman function_1}
\end{align}
Using the relation 
\[
  d^{n-1}p = \frac{2\pi^{(n-1)/2}}{\Gamma\left(\frac{n-1}{2}\right)}p^{n-2}dp,
\]
where $n$ denotes the number of spacetime dimensions, we get for a (3+1) dimension
\begin{align}
    G_+&=  \int \frac{2\pi^{3/2}}{\Gamma\left(\frac{3}{2}\right)}p^{2}dp\frac{e^{i(\bar{p}.\Delta\bar{x})}}{2\omega_p(2\pi)^{3}}
\end{align}

%Or in more general terms, we can write,
% \begin{align}
%     G_+&=  \int \frac{2\pi^{(n-1)/2}}{\Gamma\left(\frac{n-1}{2}\right)}p^{n-2}dp\frac{e^{i(\bar{p}.\Delta\bar{x})}}{2\omega_p(2\pi)^{n-1}}
% \end{align}
% \textcolor{red}{As a side note, by defining $\tilde{G}(p^2) = (p^2-m^2)^{-1}$, the Wightman function can also be rewritten as, 
% \begin{align}
%    G_+(\bar{x}, t) &=-i \int \frac{d^3\bar{p}\,dp_0}{(2\pi)^4}\tilde{G}(p^2)e^{i(\bar{p}.\Delta\bar{x}-p_0\Delta t})\label{eq: Wightman with G_tilde}
%  \end{align}}
% We refer the appendix for the proof of this equivalent notation.
Change the variables $\bar{\tau}= (\tau+\tau')/2$ and $\Delta\tau=\tau-\tau'$, the Jacobian for this transformation is one $d\tau d\tau'=d\bar{\tau}d\Delta\tau$. The response function will then be
\begin{align}
      F(\Delta E) &= \int d\bar{\tau}d\Delta\tau e^{-i(\Delta E(\Delta \tau))} G_+(\Delta\tau)
\end{align}
Considering a finite time interval from $-\t_0<\tau,\tau'<\t_0$ when the detector is interacting with the field, the response function can be factored as \begin{align}
    F(\Delta E)&=\int_{-\t_0}^{\t_0}d\bar{\tau}\int_{-\infty}^\infty d\Delta\tau e^{-i(\Delta E(\Delta \tau))} G_+(\Delta\tau)\notag\\
    &=2\tau_0\int_{-\infty}^\infty d\Delta\tau e^{-i(\Delta E(\Delta \tau))} G_+(\Delta\tau)
\end{align} 
In the limit $\tau_0\to\infty$, we obtain the response rate (per unit proper time) as
\begin{align}
    \dot{F}(\Delta E)&= \lim_{\tau_0\to\infty}\frac{F(\Delta E)}{2\tau_0}= \int d\Delta\tau e^{-i\Delta E(\Delta \tau)}  G_+(\Delta\tau)
    \label{trans_prob_rate_11}
\end{align}
Finally, since the Wightman function blows up at $\Delta\tau=0$, we regularise it with the $i\epsilon$ prescription where $G_+(\Delta\tau)\to G_+(\Delta\tau-i\epsilon)$ with $\epsilon\to 0^+$.
\begin{equation}
    \dot{F}(\Delta E)=  \int d\Delta\tau e^{-i\Delta E\Delta \tau}  G_+(\Delta\tau-i\epsilon)
\end{equation}
 Let the response rates for three processes (spontaneous, induced, and absorption) be denoted as $\dot{F}_s(E),\,\dot{F}_I(E),\, \dot{F}_A(E)$ respectively. Also, for brevity, we will use the notations $\Delta E=E$ and $\Delta \tau=\tau$ henceforth.\\

\section{Spontaneous emission rate using Rindler quantization}\label{spont_ems_1}
% From eqs. (\ref{rind_quant_1}) and (\ref{mode_rindler_1}),
The field and the rindler modes for a uniformly accelerating detector as 
\begin{equation}
\phi(x)=\int d^3 k\,\left(b_{\tilde{\omega},\bar{k}_{\perp}}\,v_{\tilde{\omega},\bar{k}_{\perp}}+\,b_{\tilde{\omega},\bar{k}_{\perp}}^{\dagger}\,v_{\tilde{\omega},\bar{k}_{\perp}}^{*}\right)\,.
\end{equation}
\begin{equation}
v_{\tilde{\omega},\bar{k}_{\perp}}=N_{\tilde{\omega}}\,K_{i \tilde{\omega}/a}\left(\frac{\sqrt{|\bar{k}_{\perp}|^2+m^2}}{a}\,e^{a\xi}\right)\,e^{-i\tilde{\omega}\tau}
\end{equation}
where $N_{\tilde{\omega}}=\sqrt{{\rm sinh}\left(\pi \tilde{\omega}/a\right)/4 \pi^4\,a}$. Here, the momentum modes for the rindler observer is $k=(\tilde{\omega}, k_\perp)$. We are also using the short-hand notation  $\int d^3 k=\int\, d\tilde{\omega}\,dk_\perp$.
% The spontaneous transition amplitude for an accelerating detector probing the vacuum state to go from $\ket{E_i,0_R}\to\ket{E_f,k}$ in rindler spacetime is
% \begin{equation}
%     A_{i\to f}=g\bra{E_f}\mu(0)\ket{E_i}\int d\tau \, e^{-i\Delta E\tau}\bra{ k}\phi(x(\tau))\ket{0_R}
% \end{equation}
% Here, $0_R$ represents the rindler vacuum. The transition probability for this spontaneous process is 
% \begin{align}
%     P&=\int d^3k\,|A_{i\to f}|^2
%     =| g\bra{E_f}\mu(0)\ket{E_i}|^2\notag\\
%     &\times \int d^3k \,d\tau \,d\tau'\e^{-i\Delta E(\tau-\tau')}\bra{ 0_R}\phi(x(\tau))\ket{k}\bra{k}\phi(x(\tau'))\ket{0_R}\notag\\
%     &= C\int \, d\tau \,d\tau' e^{-i\Delta E(\tau-\tau')}\,\bra{ 0_R}\phi(x(\tau))\phi(x(\tau'))\ket{0_R}\notag\\
%     &=C\,F_S(\Delta E) \label{eq: rindler probability_1}
% \end{align}
The response function in Minkowski quantization is 
\begin{equation}
  F_S(\Delta E)=  \int  \, d\tau \,d\tau' e^{-i\Delta E(\tau-\tau')} G_R
\end{equation} with Wightman function as 
\begin{equation}
    G_R=\bra{ 0_R}\phi(x(\tau))\phi(x(\tau'))\ket{0_R}.
\end{equation}
Since only the term containing $b_ib_j^\dagger$ will contribute a non-zero term,
\begin{align}
   G_R= \int d^3 k\, d^3k'\, \bra{0_R}(b_iv_i)(b_j^\dagger v_j^*)\ket{0_R}
\end{align}
where $i/j=(\tilde{\omega},k_\perp)$. Using the equal-time commutation relation $[b_i, b_j]=\delta^3(i-j)=\delta(\tilde{\omega}_i-\tilde{\omega}_j)\,\delta^2(k_{\perp_i}-k_{\perp_j})$, 
% \begin{align}
% &G_R=\int d^3k\,d^3k'\, v_i v_j^*\bra{0_R}\left(\delta^3(i-j)+b_j^\dagger b_i\right)\ket{0_R}\notag\\
% &=\int d^3k \,|v_i|^2\notag\\
% & = \int d^3k
% \left|N_{\tilde{\omega}}\,K_{i \tilde{\omega}/a}\left(\frac{\sqrt{|\bar{k}_{\perp}|^2+m^2}}{a}\,e^{a\xi}\right)\right|^2
% e^{-i\tilde{\omega}(\tau'-\tau)}\notag\\
%     &= \int d^3k\, \left[\frac{{\rm sinh}\left(\pi \tilde{\omega}/a\right)}{4 \pi^4\,a}\right]\left|K_{i \tilde{\omega}/a}\left(\frac{\sqrt{|\bar{k}_{\perp}|^2+m^2}}{a}\,e^{a\xi}\right)\right|^2 \notag\\
%     &\times e^{-i\tilde{\omega}(\tau'-\tau)}
% \end{align}
%%%%%
\begin{widetext}
\begin{align}
G_R &= \int d^3k\, d^3k'\, v_i v_j^{*}\,
        \bra{0_R}\left(\delta^3(i-j)+b_j^{\dagger}b_i\right)\ket{0_R} = \int d^3k\, |v_i|^2 =\int d^3k\,
        \left|N_{\tilde\omega}\,
K_{i\tilde\omega/a}\!\left(\frac{\sqrt{|\bar k_\perp|^2+m^2}}{a}\,e^{a\xi}\right)\right|^2
        e^{-i\tilde\omega(\tau'-\tau)} \notag\\
    &= \int d^3k\,
\left[\frac{\sinh\!\left(\pi\tilde\omega/a\right)}{4\pi^4 a}\right]
\left|K_{i\tilde\omega/a}\!\left(\frac{\sqrt{|\bar k_\perp|^2+m^2}}{a}\,e^{a\xi}\right)\right|^2
        e^{-i\tilde\omega(\tau'-\tau)}
\label{eq:GR}
\end{align}
Therefore the response rate becomes,
\begin{align}
\dot F_S(\Delta E) &= \int d^3k\, d(\Delta\tau)\;
        e^{i(\tilde\omega-\Delta E)(\tau-\tau')}\,
\left|K_{i\tilde\omega/a}\!\left(\frac{\sqrt{|\bar k_\perp|^2+m^2}}{a}\,e^{a\xi}\right)\right|^2
\left[\frac{\sinh\!\left(\pi\tilde\omega/a\right)}{4\pi^4 a}\right] \notag\\
    &= 2\pi \int d^3k\,
\left|K_{i\tilde\omega/a}\!\left(\frac{\sqrt{|\bar k_\perp|^2+m^2}}{a}\,e^{a\xi}\right)\right|^2
\frac{\sinh\!\left(\pi\tilde\omega/a\right)}{4\pi^4 a}\,
        \delta(\tilde\omega-\Delta E) \notag\\
    &= 2\pi \int d\tilde\omega\, dk_\perp\,
\left|K_{i\tilde\omega/a}\!\left(\frac{\sqrt{|\bar k_\perp|^2+m^2}}{a}\,e^{a\xi}\right)\right|^2
\frac{\sinh\!\left(\pi\tilde\omega/a\right)}{4\pi^4 a}\,
        \delta(\tilde\omega-\Delta E)
\label{eq:FSdot}
\end{align}
\end{widetext}
%%%%

%%% Appendix 3
\section{Verifying the periodicity}\label{periodicity_appendix}
From eq.(\ref{spt_int_1}), the spacetime interval for a uniformly accelerating detector is 
\begin{align}
  \sigma&=(t-t')^2-(x-x')^2\notag\\
  &=\frac{1}{a^2}(\sinh(a\tau)-\sinh(a\tau'))^2-\frac{1}{a^2}(\cosh(a\tau)-\cosh(a\tau'))^2\notag\\
  &=\frac{2}{a^2}[\cosh(a\Delta\tau)-1]
\end{align}
With this, the Wightman function is 
\begin{align}
G_+(\Delta\tau)&\propto \frac{1}{(\Delta t)^2-\Delta x^2}
\propto\frac{1}{\cosh(a\Delta\tau)-1}
\end{align}
Applying the Wick rotation $\Delta \tau\to\Delta \tau + 2\pi i/a$,
\begin{align}
&\cosh(a\Delta \tau+ 2\pi i)\notag\\
&= \cosh a\Delta\tau\cos(2\pi ) + i\sinh a\Delta\tau \sin(2\pi )=\cosh a\Delta\tau\notag\\
&\implies G_+(\Delta \tau + 2\pi i/a)\propto \frac{1}{\cosh(a(\Delta \tau + 2\pi i/a))-1}\notag\\
&=\frac{1}{\cosh a\Delta\tau-1}=G_+(\Delta\tau)
\end{align}

Thus, the Wightman function is periodic in the imaginary proper time with a period $2\pi/a$. Comparing this periodicity with the KMS condition for a thermal state,
\[
G_+(\Delta\tau+ i\beta)=G_+(\Delta\tau),
\]
with $\beta = 1/T$ in natural units, we can see that the periodicity in the Wightman function has given rise to a thermal state with $T = a/2\pi$.
\\
Now let's look at the response rate with $\Delta E=E$ and $\Delta\tau=\tau$,
\begin{equation}
    \dot{F}( E) = \int d\tau\,e^{-i E  \tau}G_+( \tau)
\end{equation}
By assuming $G_+$ is an analytic function in $-\beta<{\rm Im}\t<0$, we shift the contour to $\tau\to\tau-i\beta$ and use the property $G_+(\t-i\beta)=G^-(\t)=G_+(-\t)$
\begin{align}
     \dot{F}( E)&= \int d\tau\,e^{-i E ( \tau-i\beta)}G_+( \tau-i\beta)\notag\\
     &= e^{-\beta  E}\int d\t e^{-i E\t}G_+(\t-i\b)\notag\\
     &=  e^{-\beta  E}\int_{-\infty}^{\infty} d\t e^{-i E\t}G_+(-\t)
\end{align}
Letting $-\D\t=u$,
\begin{align}
      \dot{F}( E)&= -e^{-\beta  E}\int_{\infty}^{-\infty} du e^{i( E)u}G_+(u)\notag\\
      &=e^{-\beta  E}\int_{-\infty}^{\infty} du e^{-i(- E)u}G_+(u)\notag\\
   \implies   \dot{F}( E)   &= e^{-\beta E}  \dot{F}(- E)
\end{align}

%%% Appendix 4

\section{Profile function for a flat 6 dimensional model}\label{add_appendix_1}
Consider a 6 dimensional the scalar field action with flat extra dimensions givne by 
\begin{equation}
    S_6=\int d^4x \int_0^{2\pi R_1}dy\int_0^{2\pi R_2}dz \bigg[\frac{1}{2}\partial_M\Phi\partial^M\Phi-\frac{1}{2}m_0\Phi^2\bigg]
\end{equation}
where 
\begin{equation}
    \P(x^\mu, y, z)= \frac{1}{2\pi\sqrt{R_1R_2}}\sum_{n=-\infty}^{\infty}\sum_{q=-\infty}^{\infty}\phi_{nq}(x^\mu)e^{iny/R_1}e^{iqz/R_2}.
\end{equation}
For simplicity, we keep $R_1=R_2$. Integrating over $y$ and $z$, the 4-dimensional action becomes 
\begin{equation}
    S_4=\sum_{n,q=0}^\infty\int d^4x\bigg[\frac{1}{2}\partial_\mu\phi_{nq}\partial^\mu\phi_{nq}-\frac{1}{2}m_{nq}\phi_n^2\bigg]
\end{equation}
where $m_{nq}^2= m_0^2 + (n^2+q^2)/R^2$ are the mass modes. The generic form of the transition rate as given in (\ref{trans_1}), will now be modified as 
\begin{equation}
    \dot{F}( E) = \sum_{n,q=0}^\infty C \int d\tau e^{-i\tau}G_+^{(nq)}(x,x').
\end{equation}
where $m_{nm}^2=m_0^2+(n^2+q^2)/R^2$ is the KK mass modes. With this, the propagator (\ref{prop_1}) is modified as
\begin{equation}
    \tilde{G}(p)=\sum_{n=0}^\infty\sum_{m=0}^\infty\frac{1}{p^2-m_{nq}^2}.
\end{equation} 
Correspondingly, the response function is 
 \begin{equation}
    \mathcal{F}(E)
=\sum_{n=0}^\infty\sum_{m=0}^\infty\sqrt{E^2-m_{nm}^2}\Theta(E-m_{nm})
\end{equation}
and the response rate is 
\begin{equation}
    \dot{F}(E) = \frac{1}{e^{2\pi E/a}-1}\left( \sum_{n=0}^\infty\sum_{m=0}^\infty\sqrt{E^2-m_{nm}^2}\Theta(E-m_{nm})\right).
\end{equation}
 
The ADD Planck scale relation for large extra dimensions is 
\begin{equation}
    M_{Pl}^2\sim M_*^{n+2}R^n, \label{eq: gc_1}
\end{equation}
where $M_{Pl}$ is the observed Planck scale on the brane, $M_*$ is the fundamental planck scale and $n$ represents the number of extra dimensions. Fixing $M_*=1$ TeV, and $M_{Pl}=10^{16}$ TeV for $n=2$, the compactification radius of each extra dimension satisfying the above equation is of the order of $\sim \mathcal{O}(10^{16})$ TeV$^{-1}$. \\
 \\
 \subsection*{Mass tables for single and double large extra dimensions}
The following tables show first few mass modes for the flat extra-dimensional models. We have kept $R=10^{12}$ $TeV^{-1}$, for both the tables.
\begin{table}[h]
    \centering
    \begin{tabular}{|c|c|}\hline
        $n$ & $m_n \times 10^{-12}$ TeV \\ \hline
        n=0 &  0.000 \\ \hline
         n=1 & 1.000 \\ \hline
         n=2 & 2.000 \\ \hline
         n=3 & 3.000 \\ \hline 
         n=4 & 4.000 \\ \hline
    \end{tabular}
    \caption{First few mass modes for different values of $n$ for a single flat extra-dimensional model keeping $R=10^{12}$ TeV$^{-1}$. Note that all mass values are in the units of $10^{-12}$ TeV.}
    \label{tab: single flat mass modes}
\end{table}

\begin{table}[h]
\centering
\begin{tabular}{|c|c|c|c|c|}
\hline
$n \backslash q$ & $q=0$ & $q=1$ & $q=2$ & $m=3$ \\ \hline
$n=0$ & 0.000 & 1.000 & 2.000 & 3.000 \\ \hline
$n=1$ & 1.000 & 1.414 & 2.236 & 3.162 \\ \hline
$n=2$ & 2.000 & 2.236 & 2.828 & 3.606 \\ \hline
$n=3$ & 3.000 & 3.162 & 3.606 & 4.243 \\ \hline
\end{tabular}
\caption{First few mass modes for different values of $(n,q)$ for two flat extra dimensions keeping $R=10^{12}$ TeV$^{-1}$.Note that all mass values are in the units of $10^{-12}$ TeV. }
\label{tab: 6fed mass modes}
\end{table}
%%%%%%
%%% appendix 5
\section{Profile function for a 6 dimensional warped geometry model}\label{rs_appendix_1}
%%%
We consider the background spacetime metric for a 6 dimensional warped geometry model in the mostly plus  convention as follows\cite{Chakraborty:2014xda}, 
\begin{equation}                ds^2=b^2(z)\left[a^2(y)\eta_{\mu\nu}dx^\mu dx^\nu + R_y^2 \,dy^2\right] + r_z^2\,dz^2\,, \label{eq:6d_metric}
\end{equation}
where $R_y,\,r_z$ are the two radii for the compactified extra dimensions and $a(y)$, $b(z)$ represent warp factors along $y$ and $z$ directions respectively. In this model the warp factors are turn out to be as, 
\begin{align}\left.
\begin{aligned}
     b(z) = \frac{\cosh(kz)}{\cosh(k\pi)}\quad & ;\quad a(y) = e^{-c|y|}\,,\,\,\,\,\,\,\,\,{\rm where }\\
    k=r_z\sqrt{\frac{-\Lambda}{M^4}}\quad &;\quad c = \frac{R_y k}{r_z\cosh(k\pi)}\,.
\end{aligned}\right\}
\end{align}
%%%%%
The extra dimensions are orbifolded as $S^1/Z_2$. So we work on the intervals
$0\leq y,z\leq \pi$.  Therefore, 
\begin{align}       \xi_n(y)\bigg|_{0,\pi}&=0=\chi_q(z)\bigg|_{0,\pi}\notag\\
    \frac{d\xi_n}{dy}\bigg|_{0,\pi}&= 0= \frac{d\chi_q}{dz}\bigg|_{0,\pi}
\end{align}
The 6 dimensional bulk curvature is represented as, $k'=k/r_z$. 
The 6 dimensional scalar field action can be written as,  
\begin{align}
S_6=\frac{1}{2}\int d^4x\int_{0}^\pi &dy\int_{0}^\pi dz\sqrt{-g^{(6)}}\notag\\
&\times \left(-g^{MN}\partial_M\Phi\,\partial_N\Phi-m_0^2\Phi^2\right)\label{eq: 6 warped action}
\end{align}
where $m_0$ is the mass of the bulk, and 
\begin{equation}
    \Phi(x^\mu, y,z) = \sum_{n,q}\frac{\phi_{nq}(x)\,\xi_n(y)\,\chi_q(z)}{\sqrt{R_y\,r_z}}\label{kk_6d_1}.
\end{equation}
Consider the Klein-Gordon equation
\begin{equation}
    \frac{1}{\sqrt{-g}}\partial_M(\sqrt{-g}g^{MN}\partial_N\Phi)-m_0^2\Phi^2=0,
\end{equation}
Using the metric (\ref{eq:6d_metric}), the mode solution (\ref{kk_6d_1}), and requiring the 4 dimensional modes satisfy the Klein-Gordon equation, 
\begin{equation}
    (\Box-m_{nq}^2)\phi_{nq}=0,
\end{equation}
we obtain two $z$ and $y$ dependent eigen-value equations that as follows 
\begin{align}
    \frac{1}{r_z^2}\frac{d}{dz}\left(b^5\frac{d\chi_q(z)}{dz}\right)-m_0^2b^5\chi_q(z)&= -m_q^2b^3\chi_q(z)\label{eq:tower on 5}\\
    \frac{1}{R_y^2}\frac{d}{dy}\left(a^4\frac{d\xi_n(y)}{dy}\right)-m_q^2a^2\xi_n(y)&= -m_{nq}^2a^4\xi_n(y).\label{eq:tower on 4}   
\end{align}
 Here $m_q$ is the mass tower on the $y$ dimension, and $m_{nq}$ is the mass tower on the four dimensional brane. $\xi$ and $\chi$ follow the orthonormality conditions 
\begin{align}
    \int a^2(y)\xi_{n_1}(y)\xi_{n_2}(y)&=\delta_{n_1n_2}\\
    \int b^3(z)\chi_{q_1}(z)\chi_{q_2}(z)&=\delta_{q_1q_2}.
\end{align}
Using the relations (\ref{eq:tower on 4}) and (\ref{eq:tower on 5}), and applying the orthonormality and boundary conditions in (\ref{eq: 6 warped action}), gives us
 the effective 4 dimensional action as
\begin{equation}
    S_4=\frac{-1}{2}\sum_{nq}\int d^4x \left(\partial^\mu\phi_{nq}\,\,\partial_\mu\phi_{nq}+m_{nq}^2\,\phi_{nq}^2\right). \label{eq: 6d to 4d warped action}
\end{equation}
 Furthermore, solving the two differential equations, starting from the z dependent part, would gives the expression of the mass modes on the 4-D brane as
\begin{equation}
    m_{nq}= \frac{x_{\nu _n}k^\prime}{\cosh(k\pi)} \exp\left(\frac{-R_y k\pi}{r_z\cosh(k\pi)}\right)\label{eq: WED mass modes}
\end{equation}
where $x_{\nu_n}$ is the solution of the bessel function
\begin{equation}
     J_{ \nu_n}(x_{ \nu_n})+\frac{x_{ \nu_n}}{2}[J_{{ \nu_n}-1}(x_{ \nu_n})-J_{{ \nu_n}+1}(x_{ \nu_n})]=0\label{eq:bessel_6d_rs_1}.
\end{equation}
% where
% \begin{align}\left.
% \begin{aligned}
%     \nu_n&=\sqrt{1+\left(\frac{m_q}{k'}\right)^2\cosh^2(k\pi)}\\
%     \frac{m_q}{k^\prime}&= x_{\nu q}e^{-k\pi}\\
%      3J_{\nu_q}(x_{\nu_q})+&x_{\nu_q}[J_{{\nu_q}-1}(x_{\nu_q})-J_{{\nu_q}+1}(x_{\nu_q})]=0\\
%       \nu_q&=\sqrt{\frac{9}{4}+\left(\frac{M}{k^\prime}\right)^2}
% \end{aligned}\qquad \right\}
% \end{align}
Defining $R_y/r_z=\alpha$, and substituting $k^\prime=k/r_z=k\alpha/R_y$,  (\ref{eq: WED mass modes}) becomes 
\begin{equation}
     m_{nq}= \frac{x_{\nu _n}k\alpha}{R_y}\frac{1}{\cosh(k\pi)} {\rm exp}\left(\frac{- k\alpha\pi}{\cosh(k\pi)}\right)\label{eq: WED modified mass modes}.
\end{equation}
Now that we have obtained the expression of mass modes, the propagator (\ref{prop_1}) for the 6 dimensional warped extra dimension will be
\begin{equation}
    \tilde{G}(p^2) = \sum_{n,q}\frac{1}{p^2-m^2_{nq}}
\end{equation}
Correspondingly, profile function will be modified as, 
\begin{widetext}
\begin{align}
\mathcal{F}(E)&= \sum_{n,q=0}^{\infty}\sqrt{E^2-m_{nq}^2}\,\,\Theta(E-m_{nq})\notag\\
&=\mathlarger{\sum}_{n,q=0}^{\infty}\sqrt{E^2- \frac{x_{\nu _n}k^{\prime^2}}{\cosh^2(k\pi)} {\rm exp}\left[\frac{- 2R_yk\pi}{r_z\cosh(k\pi)}\right]}\,\,\Theta\left[E- \frac{x_{\nu _n}k^\prime}{\cosh(k\pi)} {\rm exp}\left(\frac{-R_y k\pi}{r_z\cosh(k\pi)}\right)\right]\notag\\
    % \label{eq: WED-2response}
&=\mathlarger{\sum}_{n,q=0}^{\infty}\sqrt{E^2- \left(\frac{x_{\nu _n}k\alpha}{R_y}\right)^2\frac{1}{\cosh^2(k\pi)} {\rm exp}\left(\frac{- 2k\alpha\pi}{\cosh(k\pi)}\right)}\,\,\Theta\left[E- \frac{x_{\nu _n} k\,\alpha}{R_y\,\cosh(k\pi)}\,{\rm exp}\left(\frac{- k\,\alpha\,\pi}{\cosh(k\pi)}\right)\right]\label{eq: WED-2 modified response}.
\end{align}
\end{widetext}
and the response rate is 
\begin{equation}
    \dot{F}(\D E) = \frac{1}{e^{2\pi\, \D E/a}-1}\left[\sum_{n,q=0}^{\infty}\sqrt{E^2-m_{nq}^2}\,\,\Theta(E-m_{nq})\right].
\end{equation}
For the 6 dimensional RS scenario we fix $k=0.25$. Resolution of hierarchy problem demand that, 
\begin{equation}
    \frac{e^{-c\pi}}{\cosh(k\pi)}\sim10^{-16}.
\end{equation}
This equation fixes the value of $\alpha=61.66$. 
 \subsection*{Mass tables for single and doubly warped geometry models}
 The following tables show the first  mass modes for the flat extra dimensional models. 
\begin{center}
    \centering
    \begin{tabular}{|c|c|c|}\hline
  $n$ &  $m_n$ (TeV) & $\Delta m_n$ \\\hline
$n=0$ & 0  & 0 \\ \hline
$n=1$ & 7.8174& 7.8174 \\ \hline
$n=2$ & 14.3132& 6.4992  \\ \hline
$n=3$ & 20.7559&6.4427 \\ \hline
$n=4$ & 27.1829& 6.4270  \\ \hline
$n=5$ & 33.6033 & 6.4204  \\ \hline
\end{tabular}~~
\begin{tabular}{|c|c|c|} \hline
    $n$ & $m_n$ & $\Delta m_n$  \\ \hline
    $n=6$ & 40.0202 & 6.4169 \\ \hline
     $n=7$ & 46.350 & 6.4148\\ \hline
    $ n = 8$ & 52.846 & 6.4136 \\ \hline
      $n=9$ & 59.2612 & 6.4126 \\ \hline
    $n=10$ & 65.6733 & 6.4121 \\ \hline
    $n= 11$ & 72.0848 & 6.4115 \\ \hline
\end{tabular}
\captionof{table}{First few mass modes for different values of $n$ for a single warped extra dimension keeping $k = 10^{16}$ TeV, and $R_y = 1.15\times 10^{-15}$ TeV$^{-1}$ }

    \label{tab:5wed mass modes}
\end{center}
%%%
% In order to understand why the nature of the profile function with respect to the $R_y$, let us consider the mass modes for different values of $R_y$. For a single warped extra dimension, we had taken $R_y = 1.15\times 10^{-15}$ TeV$^{-1}$. Let us see the effect of different values of $R_y= 1\times 10^{-15}, 1.15\times 10^{-15}, 1.5\times 10^{-15}, 2
% \times 10^{-15}$ on the mass modes.

\begin{center}
    \centering
     \resizebox{\columnwidth}{!}{
    \begin{tabular}{|c|c|c|c|}\hline
     $R_y= 1\times10^{-15} $ & $R_y= 1.15\times 10^{-15}$    &$R_y= 1.5\times 10^{-15}$&$R_y=  2
\times 10^{-15}$ \\\hline
     0 & 0  & 0 & 0\\\hline 
     870.22 & 7.8174 & $1\times10^{4}$ &$1.97\times10^{-11}$ \\\hline
     1593.31 & 14.3132 & $2\times10^{4}$  &$3.61\times10^{-11}$\\\hline
     2310.49 & 20.7559 & $3\times10^{4}$ & $5.24\times10^{-11}$ \\\hline
     3025.94 & 27.1829 &$5\times10^{4}$ & $6.8722\times10^{-11}$ \\\hline
     3740.64 & 33.6033 & $6\times10^{4}$ & $8.495\times10^{-11}$\\ \hline
    \end{tabular}}
    \captionof{table}
    {Mass modes (in TeV) for different values of $R_y$ (TeV$^{-1}$) for a single warped extra dimension. Here $k=10^{16}
    $ TeV.}
    \label{tab:placeholder}
\end{center}
%%%%%
\begin{center}
\centering
% \resizebox{columnwidth}{!}{
\begin{tabular}{|c|c|c|c|c|}
\hline
$n \backslash q$ & $q=0$ & $q=1$ & $q=2$ & $q=3$ \\\hline
$n=0$ & 0.567 & 0.865 & 1.171 & 1.478 \\ \hline
$n=1$ & 1.088 & 1.437 & 1.786 & 2.129 \\ \hline
$n=2$ & 1.584 & 1.9530 & 2.322 & 2.685 \\ \hline
$n=3$ & 2.074 & 2.4552 & 2.837 & 3.212 \\ \hline
\end{tabular}
\captionof{table}{Mass modes for different values of $n$ and $q$ for doubly warped extra dimensions keeping $k = 0.25$,  $\alpha = 61.6605$, and  $R_y = 10^{-14}$ TeV$^{-1}$. All the masses are in TeV.   }
\label{tab: 6wed mass modes}
\end{center}
\vspace{50cm}
\bibliographystyle{apsrev4-2}
\bibliography{biblography}
%%%%%%%%%%%%%%%%%%%%%%%%%%%%%%%%%%%%%%%%
\end{document}